\documentclass[prl, reprint, twocolumn, aps, superscriptaddress]{revtex4-2}
\usepackage{graphicx}
\usepackage{dcolumn}
\usepackage{bm}
\usepackage{dcolumn} % Align table columns on decimal point
\usepackage{color,soul}
\definecolor{purple}{RGB}{127,0, 255}

\usepackage{ragged2e}
\usepackage{amssymb}
\usepackage{amsmath}
\usepackage{soul}

\begin{document}

%%%%%%%%%%%%%%%%%%%%%%%%%%%%%%%%%%%%%%%%%%%%%%%%%%%%%%%%%%%%%%%%%%%%%%%%%%%
\title{Two-Fold Anisotropic Superconductivity in Bilayer $T_d$-MoTe$_2$}
%%%%%%%%%%%%%%%%%%%%%%%%%%%%%%%%%%%%%%%%%%%%%%%%%%%%%%%%%%%%%%%%%%%%%%%%%%%
%%%%%%%%%%%%%%%%%%%%%%%%%%%%%%%%%%%%%%%%%%%%%%%%%%%%%%%%%%%%%%%%%%%%%%%%%%%
% AUTHORS
\author{Zizhong Li}
\thanks{These two authors contributed equally}
\affiliation{Department of Materials Science and Engineering, University of Wisconsin-Madison, Madison, WI, 53706}
\author{Apoorv Jindal}
\thanks{These two authors contributed equally}
\affiliation{Department of Physics, Columbia University, New York, NY, 10027}
\author{Alex Strasser}
\affiliation{Department of Materials Science and Engineering, Texas A$\&$M University, College Station, TX, 77843}
\author{Yangchen He}
\affiliation{Department of Materials Science and Engineering, University of Wisconsin-Madison, Madison, WI, 53706}
\author{Wenkai Zheng}
\affiliation{Department of Physics, Florida State University, Tallahassee, FL, 32306}
\affiliation{National High Magnetic Field Laboratory, Tallahassee, FL, 32310}
\author{David Graf}
\affiliation{National High Magnetic Field Laboratory, Tallahassee, FL, 32310}
\author{Takashi Taniguchi}
\affiliation{National Institute for Materials Science, Tsukuba, Japan}
\author{Kenji Watanabe}
\affiliation{National Institute for Materials Science, Tsukuba, Japan}
\author{Luis Balicas}
\affiliation{National High Magnetic Field Laboratory, Tallahassee, FL, 32310}
\author{Cory R. Dean}
\affiliation{Department of Physics, Columbia University, New York, NY, 10027}
\author{Xiaofeng Qian}
\email{feng@tamu.edu}
\affiliation{Department of Materials Science and Engineering, Texas A$\&$M University, College Station, TX, 77843}
\author{Abhay N. Pasupathy}
\email{apn2108@columbia.edu}
\affiliation{Department of Physics, Columbia University, New York, NY, 10027}
\author{Daniel A. Rhodes}
\email{darhodes@wisc.edu}
\affiliation{Department of Materials Science and Engineering, University of Wisconsin-Madison, Madison, WI, 53706}

\begin{abstract}
Noncentrosymmetric 2D superconductors with large spin-orbit coupling offer an opportunity to explore superconducting behaviors far beyond the Pauli limit. One such superconductor, few-layer $T_d$-MoTe$_2$, has large upper critical fields that can exceed the Pauli limit by up to 600\%. However, the mechanisms governing this enhancement are still under debate, with theory pointing towards either spin-orbit parity coupling or tilted Ising spin-orbit coupling. Moreover, ferroelectricity concomitant with superconductivity has been recently observed in the bilayer, where strong changes to superconductivity can be observed throughout the ferroelectric transition pathway. Here, we report the superconducting behavior of bilayer $T_d$-MoTe$ _2$ under an in-plane magnetic field, while systematically varying magnetic field angle and out-of-plane electric field strength. We find that superconductivity in bilayer MoTe$_2$ exhibits a two-fold symmetry with an upper critical field maxima occurring along the $b$-axis and minima along the $a$-axis. The two-fold rotational symmetry remains robust throughout the entire superconducting region and ferroelectric hysteresis loop. Our experimental observations of the spin-orbit coupling strength (up to 16.4 meV) agree with the spin texture and spin splitting from first-principles calculations, indicating that tilted Ising spin-orbit coupling is the dominant underlying mechanism. 
\end{abstract}
\maketitle
\begin{figure*}[t]
    \includegraphics[width=0.95\linewidth]{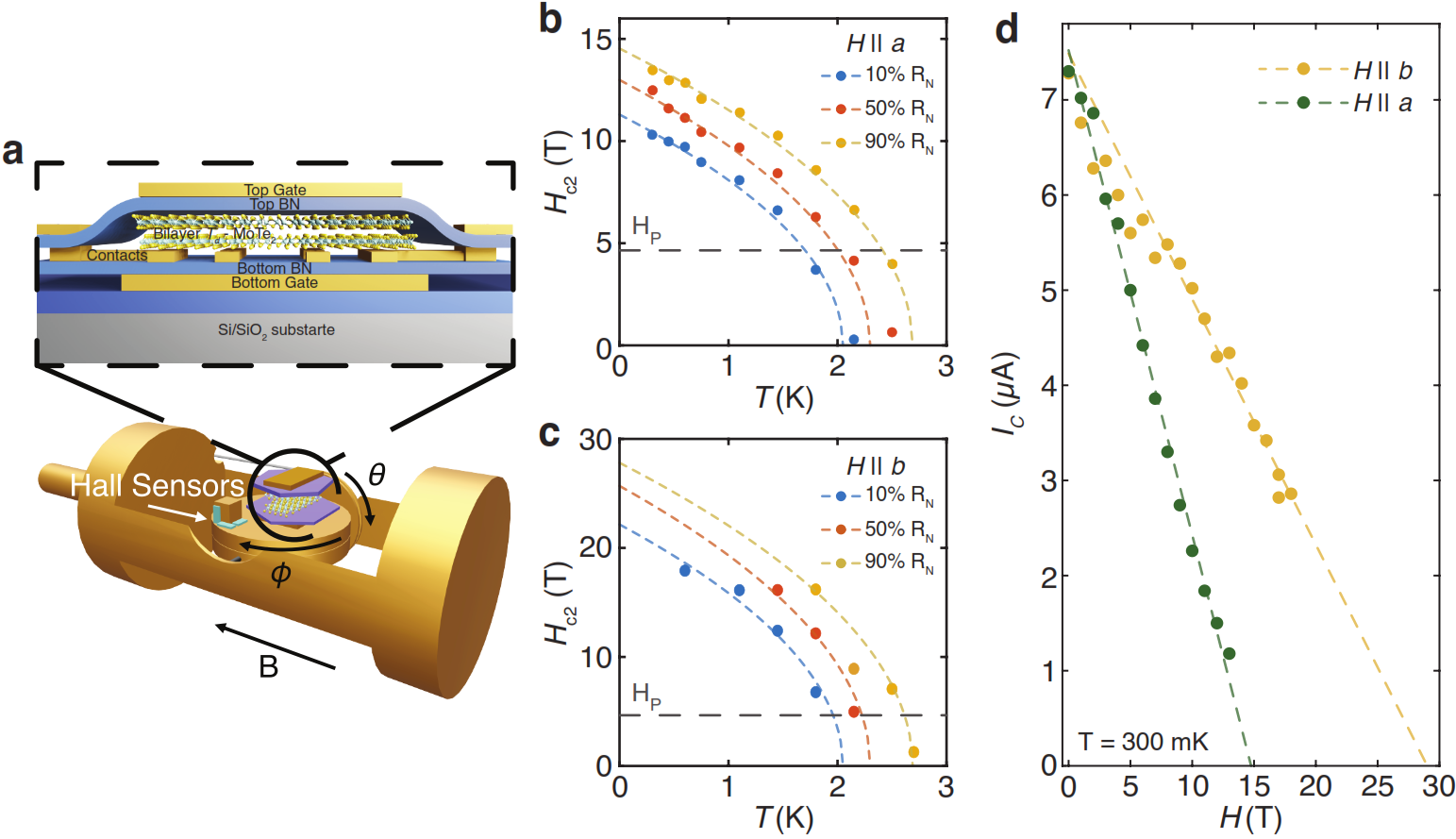}
    \caption{(a) Schematic for the two-axis rotator stage and bilayer $T_d$-MoTe$_2$ device heterostructure. (b, c) $H_{c2}$ versus $T$ for magnetic fields parallel to the $a$ (b) and $b$ (c) axes. Blue, orange, and yellow dashed lines denote $H_\textrm{c2}$ values extracted at 10\%, 50\%, and 90\%$R_\textrm{N}$, respectively. (d) Magnetic field dependence of the critical current for fields parallel to the $a$ and $b$ axes.}
\label{F2}
\end{figure*}

%%%%%%%%%%%%%%%%%%% Introduction %%%%%%%%%%%%%%%%%%%%% 
In recent years, many two-dimensional (2D) superconductors (SCs) have been shown to substantially exceed the Pauli limit ($H_\textrm{p}$). However, the primary mechanisms governing these enhancements are difficult to assign, as many different mechanisms are possible, including Ising spin-orbit coupling (SOC) types I (intervalley~\cite{lu2015evidence,xi2016ising,de2018tuning}) and II (interorbital~\cite{wang2019type,falson2020type,liu2020type}), tilted Ising SOC~\cite{cui2019transport,rhodes2021enhanced}, dynamic spin-momentum locking~\cite{yoshizawa2021atomic}, as well as  the recently proposed spin-orbit parity coupling (SOPC)~\cite{xie2020spin,zhang2023spin}. For few-layer MoTe$_2$, evidence for either SOPC\cite{xie2020spin} or tilted Ising SOC has been reported~\cite{cui2019transport,rhodes2021enhanced}. Both mechanisms can enhance the in-plane upper critical fields up to multiple values of $H_\textrm{p}$ and follow the same Ginzburg-Landau square root-dependence on temperature. However, lack of accurate in-plane angle-dependent measurements on pristine samples with a nearly intrinsic Fermi level leave open the question as to which mechanism is dominant. For SOPC to lock spins and enhance the upper critical field there must exist strong orbital pair mixing near the Fermi level. In few-layer WTe$_2$ and MoTe$_2$, this is realized through band inversion, which also enables topological edge states~\cite{qian2014quantum,tang2017quantum}. For tilted Ising SOC, spins are locked by strong spin-orbit coupling, enabled by broken inversion symmetry. Without disorder, the enhancement of upper critical fields via SOPC (max $\sim$2.5$H_\textrm{p}$ for the 1T$^\prime$ structure) is generally weaker than the enhancement by tilted Ising SOC. As a result, the differences between the two mechanisms can be discerned by mapping the enhancement of $H_\textrm{c2}$ and its anisotropy along in-plane directions. Understanding the underlying mechanism responsible for enhanced upper critical fields is important for exploring unconventional states in 2D SCs. This is particularly true for exploring finite momentum states~\cite{sato2017topological,lee2021gate,liu2021discovery,wei2023discovery,wan2023orbital} that exceed $H_\textrm{p}$ and establish a Fulde-Ferrell-Larkin-Ovchinnikov (FFLO) state and for determining potential candidates for topological superconductivity. In this Letter we provide insight on the mechanism that governs the enhanced upper critical fields and the nontrivial superconducting properties of bilayer $T_d$-MoTe$_2$ by examining the interplay between SOC and superconductivity. This is achieved by a systematic exploration of the effects of applied in-plane magnetic and out-of-plane electric fields through precise alignment of bilayer $T_d$-MoTe$_2$ using a unique two-axis, piezo-rotator and spring combination.\\ 
%%%%%%%%%%%%%%%%%%%%%%%% Begin FIGURE %%%%%%%%%%%%%%%%%%%%%%%%%%%%%%%%
\begin{figure*}[t]
    \includegraphics[width=0.7\linewidth]{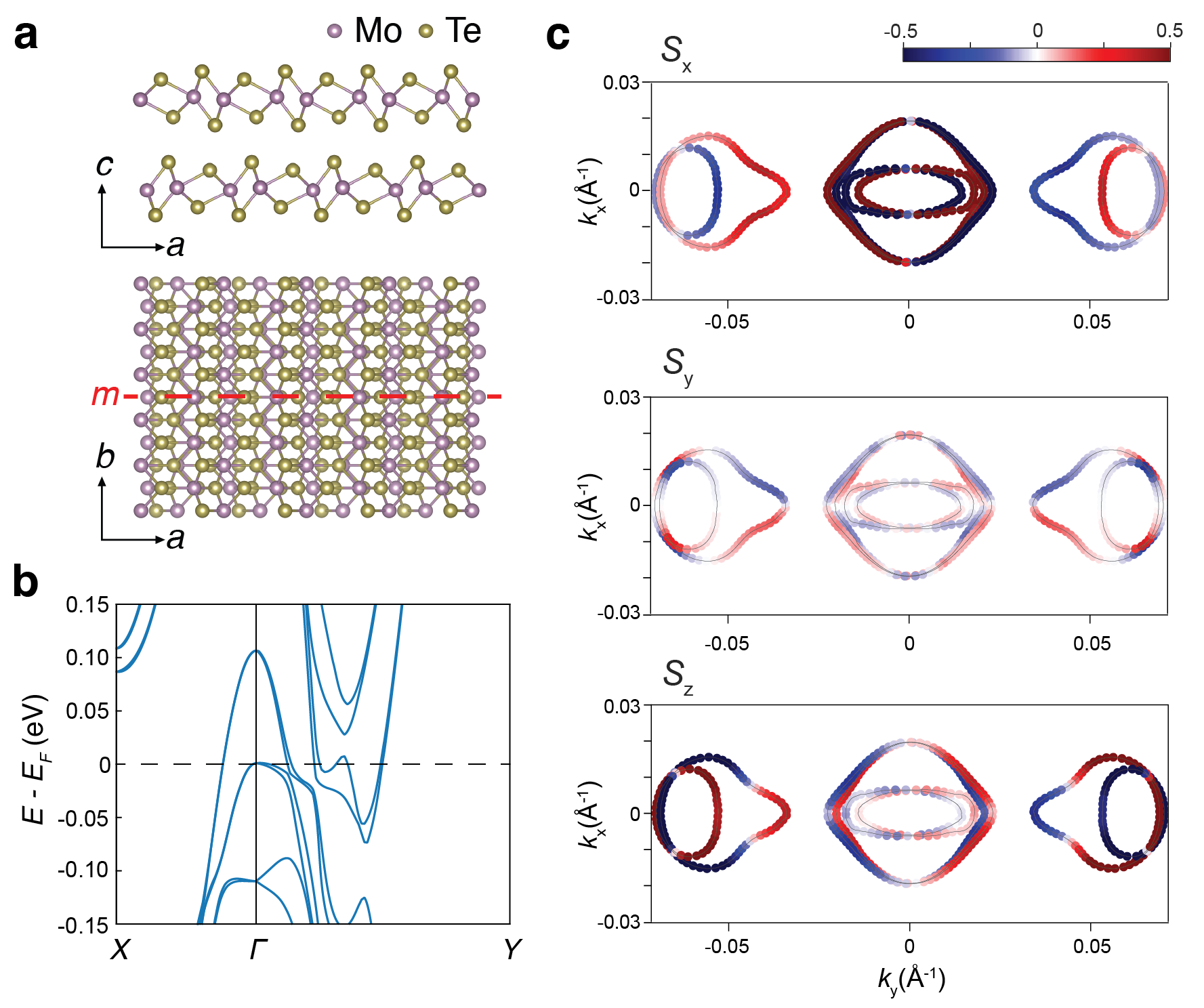}
    \caption{(a) The side view (top) and the top view (bottom) crystal structure of bilayer MoTe$_2$. The mirror plane is labeled as the dashed red line. (b) Electronic band structure for bilayer MoTe$_2$ calculated by DFT with spin-orbit coupling included. (c) Spin texture projection for bilayer MoTe$_2$ close to the Fermi level with $E=E_F - 0.01$ eV for the $S_x$, $S_y$, and $S_z$ components.}
\label{F1}
\end{figure*}
%%%%%%%%%%%%%%%%%%%%%%%%%%%%%%%%%%%%%%%%%%%%%%%%%%%%%%%%%%%%%%%%%%%%.
%%%%%%%%%%%%%%%%%%%%%%%%%%%% End Intro %%%%%%%%%%%%%%%%%%%%%%%%
Bilayer $T_d$-MoTe$_2$ has a noncentrosymmetric, orthorhombic crystal structure with Mo-chains along the $b$-axis and a mirror plane symmetry along the $a$ and $c$ axes, as shown in Fig. \ref{F1}a. As a result of the mirror symmetry, significant SOC is expected with a two-fold rotational symmetry along the in-plane direction. From the density functional theory (DFT) calculated electronic band structure (Fig. \ref{F1}b), we observe two spin-split electron-like bands and two spin-split hole-like bands crossing the Fermi level, with the hole bands centered around the $\Gamma$ point. Consistent with both monolayer and bulk MoTe$_2$, the DFT-calculated carrier density indicates a nearly-perfectly compensated semimetal with hole ($n_h$) and electron ($n_e$) densities equal to $\sim2.2\times10^{13}$ cm$^{-2}$ at the Fermi level. These values are in good agreement with the carrier densities ($n_e=2.2\times10^{13}$ cm$^{-2}$, $n_h = 2.6\times10^{13}$ cm$^{-2}$ without gating) extracted from Hall measurements using a two-band semiclassical model~\cite{jindal2023coupled}. The combination of broken inversion symmetry and large SOC gives rise to two distinct spin textures in the bilayer (see Fig. \ref{F1}c). For the hole pocket centered at $\Gamma$, the spin texture is nontrivial with low SOC strength and oscillates in- and out-of-plane along $k_x$ and $k_y$. For the electron pockets flanking $\Gamma$ (akin to the $\pm$Q pockets for monolayer MoTe$_2$~\cite{rhodes2021enhanced}), the spin texture has large SOC strength and resides primarily out-of-plane. Given the mirror-plane symmetry, anisotropic spin textures calculated by DFT, and prior models comparing different directions of the in-plane spin susceptibility~\cite{cui2019transport}, we expect that bilayer MoTe$_2$ will exhibit anisotropic superconducting properties under varying in-plane and out-of-plane magnetic field angles.\\
\indent To systematically study the effects of the spin texture on the superconducting state, we fabricated a dual-gated heterostructure of bilayer $T_d$-MoTe$_2$ encapsulated in hexagonal boron nitride (hBN), allowing for continuous tuning of doping and out-of-plane electric field (see Supplemental Material for details). Due to the large magnetoresistance~\cite{chen2016extremely} for out-of-plane magnetic fields, the possible tilted Ising superconductivity~\cite{rhodes2021enhanced,cui2019transport} in MoTe$_2$, and the cusp-like nature of critical fields (from in- to out-of-plane) for 2D SCs~\cite{tinkham2004introduction,saito2016highly}, small cants away from parallel in-plane field directions can lead to large changes in the sample resistance and obfuscate the intrinsic in-plane rotational symmetry~\cite{hamill2021two,wan2023orbital}. To ensure that we are perfectly aligned in-plane, we use a two-axis rotator stage (Fig. \ref{F2}a). The rotation angle of the sample along the polar direction, $\theta$, is controlled by a spring rotator, while the azimuthal angle, $\phi$, is controlled by a full 360$^{\circ}$ piezo-rotator. By applying voltage to the top ($V_\textrm{tg}$) and bottom gates ($V_\textrm{bg}$) we can simultaneously and independently control the carrier density and applied displacement field, $D$. Because MoTe$_2$ is semimetallic, we indicate changes to the carrier density as $\Delta n$, rather than an absolute density. Details for the calculated values of $\Delta n$ and $D$ are outlined in the Supplemental Material. For all measurements, current is passed along the $b$-axis, which we identify as the long axis of a cleaved flake~\cite{beams2016characterization}.
%%%%%%%%%%%%%%%%%%%%%%%%%%%%%%%%%%%%%%%%%%%
Using this setup, we first measure the in-plane upper critical fields along the $a$ and $b$ axes, $H_\textrm{c2}^a$ and $H_\textrm{c2}^b$, respectively (Fig. \ref{F2}b,c). As expected, we observe a clear difference in the maximum $H_{c2}$ between the two crystallographic directions due to the anisotropic behavior of MoTe$_2$ when fields are aligned in-plane~\cite{cui2019transport}. We analyze the overall trend along each direction using the thin film pair-breaking equation~\cite{tinkham2004introduction,xi2016ising}:
%%%%%%%%%%%%%%%%%%%%%%%%%%%%%%%%%%%%%%%%%%%%
\begin{center}
$\ln(\frac{T_\textrm{c}}{T_\textrm{c0}})+\psi(\frac{1}{2}+\frac{\mu_\textrm{B}H^2_\parallel/H_\textrm{p}}{2\pi k_\textrm{B}T_\textrm{c}})-\psi(\frac{1}{2})=0$,\\
\end{center}
%%%%%%%%%%%%%%%%%%%%%%%%%%%%%%%%%%%%%%%%%%%%
%%%%%%%%%%%%%%%%%%%%%%%% Begin FIGURE 2 %%%%%%%%%%%%%%%%%%%%%%%%%%%%%%%%
%%%%%%%%%%%%%%%%%%%%%%%%%
%%%%%%%%%%%%%%%%%%%%%%%%%%%%%%%%%%%%%%%%%%%%%%%%%%%%%%%%%%%%%%%%%%%
%%%%%%%%%%%%%%%%%%%%%%%%%%%%%%%%%%%%% Figure 3 %%%%%%%%%%%%%%%%%%%%%%%%%%%%%%%%%%
\begin{figure*}[t]
    \includegraphics[width=0.95\linewidth]{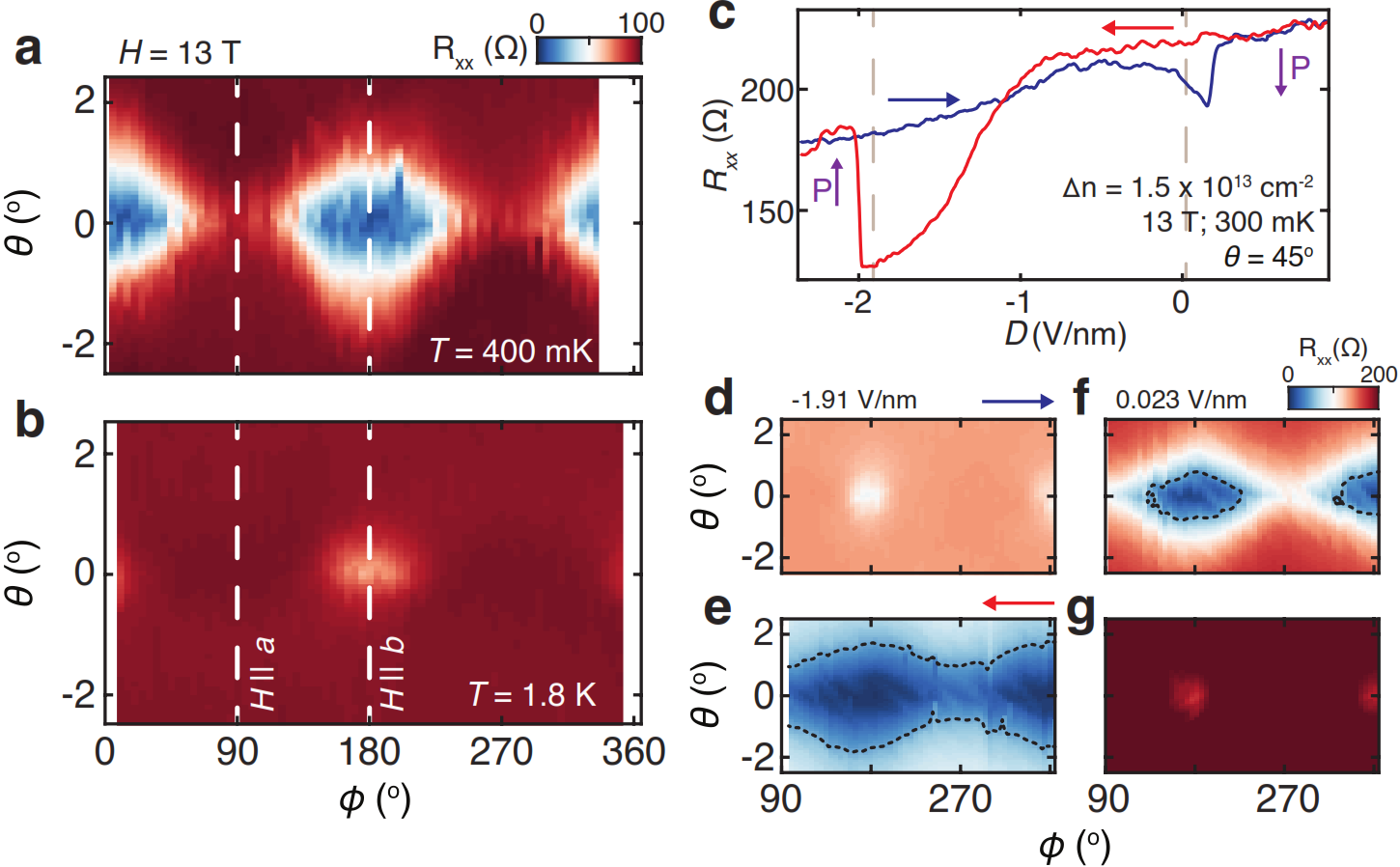}
    \caption{In-plane symmetry and ferroelectricity. (a, b) Mapping of rotation angle dependent resistance at 350 mK, 13 T (a) and 1.8 K, 13 T (b), indicating a two-fold anisotropy within the whole superconducting range. Dashed white lines denote the positions for magnetic field parallel to the $a$ and $b$ axes. (c) Ferroelectric switching of bilayer MoTe$_2$ with varying  displacement field, $D$. The internal polarization is marked with purple arrows. Red and blue arrows indicate the sweeping directions for $D$. (d-g), Evolution of superconductivity with varying $D$ along the ferroelectric hysteresis loop for $T = 300$ mK and $B = 13$ T. The chosen values of $D$ are indicated in (c) with dashed lines. The dashed black lines in (e, f) denote the superconducting regions for the 50\%$R_\textrm{N}$ criteria.}
\label{F3}
\end{figure*}
%%%%%%%%%%%%%%%%%%%%%%%%%%%%%%%%%%%%%%%%%%%%%%%%%%%%%%%%%%%%%%%%%%%%%%%%%%%%%%%%%%
%%%%%%%%%%%%%%%%%%%%%%%%%%%%%%%%%%%%% Figure 4 %%%%%%%%%%%%%%%%%%%%%%%%%%%%%%%%%%
\begin{figure*}[t]
    \includegraphics[width=0.75\linewidth]{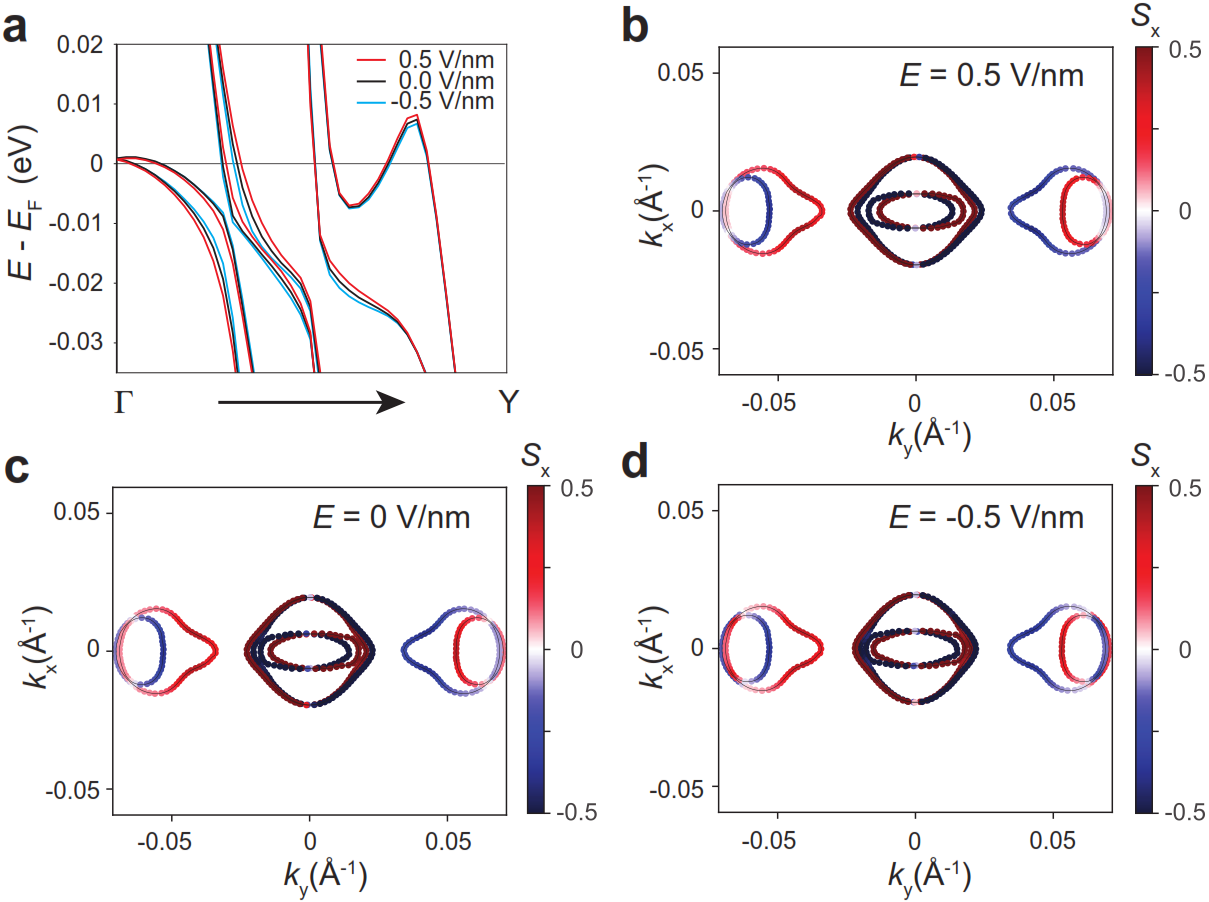}
    \caption{(a) Electronic band structure of bilayer MoTe$_2$ under varying out-of-plane electric field. (b, c, d) Spin texture projection of bilayer MoTe$_2$ at the Fermi level for $S_x$ component under the out-of-plane electric field of $E_z=0.5, 0, -0.5$ V/nm, respectively.}
\label{F4}
\end{figure*}
%%%%%%%%%%%%%%%%%%%%%%%%%%%%%%%%%%%%%%%%%%%%%%%%%%%%%%%%%%%%%%%%%%%%%%%%%%%%%%%%%
where $\psi(x)$ is the digamma function, $k_\textrm{B}$ is the Boltzmann constant, and $\mu_\textrm{B}$ is the Bohr magneton, from which we can estimate $H_\textrm{c2}^a(0)$ and $H_\textrm{c2}^b(0)$. Together, $\mu_BH^2_\parallel/H_\textrm{p}$ represents the effective pair-breaking energy for the superconductor, equivalent to an effective Zeeman energy brought about by spin-splitting~\cite{sigrist2009introduction,youn2012role}. Near $T_\textrm{c0}$, this equation can be reduced to $H_\textrm{c2} = \sqrt{H_\textrm{so}H_\textrm{p}(1-T_\textrm{c}/T_\textrm{c0})}$, where $H_\textrm{so}^\perp$ represents an effective out-of-plane magnetic field as a result of the SOC. Using this reduced equation and keeping $H_\textrm{so}^\perp$ as a free parameter, we fit the temperature-dependent $H_\textrm{c2}$ data for fields along both the $a$- and $b$-axes (Fig. \ref{F2}b,c). By extrapolating using $H_\textrm{c2}$ values at 50\% of the zero-field normal state resistance ($R_\textrm{N})$, we find $H^{\parallel a}_\textrm{c2}(0)$ and $H^{\parallel b}_\textrm{c2}(0)$ are equal to 13 T and 25.7 T, respectively. To further verify the accuracy of these fittings, we also measured critical current ($I_\textrm{c}$) as a function of in-plane magnetic field. As shown in Fig. \ref{F2}d, we observe a linear-dependence for both axes, in agreement with the expectations of mean-field theory~\cite{galitski2001disorder}. The zero-temperature upper critical fields extracted from the $I_\textrm{c}$ data are $H_\textrm{c2}^{\parallel b}(0)=28.9$ T, $H_\textrm{c2}^{\parallel a}(0)=14.7$ T, consistent with the values extracted from the temperature-dependent critical field data. We note that the anisotropy of the superconducting behavior persists for $I_\textrm{c}$, suggesting an anisotropic response of the superconducting gap with in-plane magnetic field, similar to 2M-WS$_2$~\cite{zhang2023spin}.\\
\indent Both $H_\textrm{c2}^{\parallel a}(0 \textrm{K})$ and $H_\textrm{c2}^{\parallel b}(0 \textrm{K})$ are well above the Pauli limit ($H_\textrm{p}=1.84T_\textrm{c}=4.6$ T), with $H^{\parallel a}_\textrm{c2}(0 \textrm{K})= 2.9H_\textrm{p}$ and $H^{\parallel b}_\textrm{c2}(0 \textrm{K})= 5.6H_\textrm{p}$. The amount to which $H^{\parallel b}_\textrm{c2}(0 \textrm{K})$ exceeds the Pauli limit in bilayer MoTe$_2$ is far beyond the values observed in exfoliated monolayers (1.9$H_\textrm{p}$)~\cite{rhodes2021enhanced} and chemical vapor deposition (CVD) grown few-layer flakes (2.8$H_\textrm{p}$) of MoTe$_2$. The anisotropy between the $a$ and $b$ axes ($H^{\parallel b}_\textrm{c2}(0 \textrm{K})/H^{\parallel a}_\textrm{c2}(0 \textrm{K})$) is 1.98, a substantial increase as compared to few-layer, CVD-grown MoTe$_2$ (1.53)~\cite{cui2019transport}, and bulk single crystals (1.2)~\cite{rhodes2017bulk}. Given this large anisotropy and the additional observation of Shubnikov-de Haas oscillations in the normal state~\cite{jindal2023coupled}, it is unlikely that the enhancement in $H_\textrm{c2}$ is the result of spin-orbit scattering~\cite{klemm1975theory}. In previous reports~\cite{rhodes2021enhanced,cui2019transport}, the enhanced $H^\parallel_\textrm{c2}$ combined with the in-plane anisotropy in few-layer MoTe$_2$ was initially considered to result from strong SOC combined with a complex spin texture inherent to the broken inversion symmetry. In this scenario, the majority of spin was considered to be locked along the $x-z$ plane, leading to enhanced upper critical fields along the $y$ direction. As stated earlier, SOPC has also been suggested as the mechanism for enhancing $H^\parallel_\textrm{c2}$ in MoTe$_2$. When combined with disorder, SOPC predicts $H^\parallel_\textrm{c2}$'s of up to 4$H_\textrm{p}$ along the $b$-axis with anisotropic in-plane behavior and no SOPC enhancement of the upper critical field along the $a$-axis~\cite{xie2020spin}.\\ 
\indent Since SOPC can influence superconductivity in both centro- and noncentrosymmetric superconductors, recent works have ruled out Ising-type mechanisms by focusing on centrosymmetric 1T$^\prime$ and 2M superconductors. The symmetry of multilayer MoTe$_2$ can be difficult to determine, as Raman studies have shown that both the $T_d$ (noncentrosymmetric) and 1$T^\prime$ (centrosymmetric) structures can exist at low temperature~\cite{cheon2021structural}. In addition, transmission electron microscopy studies have shown significant interlayer stacking disorder for bulk and few-layer mechanically exfoliated flakes~\cite{hart2023emergent}. Prior results also found that $H^{\parallel}_\textrm{c2}$ for the $b$-axis remains below 4$H_\textrm{p}$ with no other evidence for broken inversion symmetry beyond the enhanced $H^\parallel_\textrm{c2}$ in few-layer MoTe$_2$ and so could not rule out SOPC~\cite{cui2019transport,rhodes2021enhanced}. For bilayer MoTe$_2$, due to the flipped orientation between the two layers, both $T_d$ and 1$T^\prime$ and any stacking disorder via shear between layers will still break inversion symmetry, allowing tilted Ising SOC to contribute to the enhanced in-plane upper critical fields. In transport, the broken inversion symmetry of bilayer MoTe$_2$ is evidenced by the existence a ferroelectric transition~\cite{jindal2023coupled}.\\
\indent In contrast to prior works, in our bilayer MoTe$_2$ devices $H^\parallel_\textrm{c2}$ along the $b$-axis far exceeds 4$H_\textrm{p}$. For the $a$-axis, $H^\parallel_\textrm{c2}$ significantly surpasses 1.2$H_\textrm{p}$ (the maximum calculated value in the clean limit for SOPC). In combination with the anisotropy, our data is in strong agreement with the tilted Ising SOC model with spins locked along the $x-z$ plane. To further confirm this, we extract the SOC strength ($\Delta_\textrm{so}$) for the out-of-plane component from the pair-breaking equation, giving values of $2\Delta_{so}^{\|b}$ = 16.4 meV for the $b$-axis and $2\Delta_{so}^{\|a}$ = 4.3 meV for the $a$-axis. These values are in general agreement for the expected spin-splitting of the conduction band near the Fermi level (Fig. \ref{F1}b). Altogether, these results suggest that tilted Ising SOC is the dominant mechanism for the observed enhanced and anisotropic upper critical fields.\\
\indent To further explore the in-plane symmetry of the superconducting behavior in bilayer MoTe$_2$, we perform magnetotransport measurements while simultaneously rotating the field in- ($\theta$ and) out-of-plane ($\phi$). In Fig. \ref{F3}, we map $R_{xx}$ as a function of both $\theta$ and $\phi$ with the backlash from the string rotator subtracted out (see Supplemental Material). Here, $\theta=0$ represents complete in-plane alignment and $\phi=0^\circ$ and $90^\circ$ represent fields aligned along the $b$ and $a$ axes, respectively. As shown in Fig. \ref{F3}a, we begin by mapping the anisotropy of $H_\textrm{c2}^\parallel$ at 350 mK with a constant applied magnetic field of 13 T, close to the value of $H_\textrm{c2}^{\parallel a}(0 \textrm{K})$. We observe a clear two-fold rotational symmetry for fields along the in-plane direction, where superconductivity persists for fields aligned along the $b$-axis and the normal state resistance is reached for fields along the $a$-axis. We note that there is no anisotropy for in-plane field direction observed in the normal state magnetoresistance ($\theta>2.5^\circ$), indicating that this effect is intrinsic to the superconducting state. This two-fold rotational symmetry is consistent across bilayer samples (see Supplemental Material) and does not change for all measured temperatures up to 1.8 K ($T/T_\textrm{c0}=0.7$, Fig. \ref{F3}b). These results are contrary to those that have been seen in other 2D superconductors, like NbSe$_2$, where two-fold and six-fold rotational symmetries depend on temperature and field. The former has been suggested as being due to $s$-wave and $d$-wave pair mixing, enabled by strain~\cite{hamill2021two,haim2022mechanisms}, and the latter by a transition to a FFLO state~\cite{wan2023orbital}.\\
\indent Given our recent report of ferroelectric behavior coupled to superconductivity~\cite{jindal2023coupled}, it is natural to wonder whether there could be any changes to the two-fold rotational symmetry as bilayer MoTe$_2$ transitions from a ferroelectric to a paraelectric state along the interlayer sliding pathway~\cite{yang2018origin,liu2019vertical}. In Fig. \ref{F3}c, we confirm a ferroelectric response with respect to displacement field ($D$). Superconductivity maximizes just before the switching of polarization, consistent with our earlier report~\cite{jindal2023coupled}. Here, we have chosen to dope the sample with $\Delta n = 1.5\times10^{13}$ cm$^{-2}$ electrons to maximize the hysteretic behavior. As shown in Fig. \ref{F3}d-f, we perform the same mapping of $R_{xx}$ versus $\theta$ and $\phi$ with field and temperature identical to that of Fig. \ref{F3}a, but with $D$ set to values before and after the switching of polarization for forward and reverse directions. To ensure that we have completely switched to the ``up" polarization state, we first bias the device with $D=-2.32$ V/nm and then proceed to map $R_{xx}$ versus $\theta$ and $\phi$ before and after switching to the ``down" polarization state for the same two values of $D$, -1.91 V/nm and 0.023 V/nm. We choose these values due to their close proximity to the voltages required for switching the internal polarization and the voltages required for maximizing $T_\textrm{c}$ (-1.91 V/nm for ``down" and 0.023 V/nm for ``up," see the dashed lines in Fig. \ref{F3}c). We observe clear hysteretic behavior in these maps, reflected by the changing $R_\textrm{N}$ and $H_\textrm{c2}$ for the same values of $D$. However, much like in the $D = 0$ V/nm case discussed earlier, we observe the same two-fold rotational symmetry for the superconducting state. From these results, and combined with the fact that we observe ferroelectric behavior, we can conclude that inversion symmetry is broken in bilayer MoTe$_2$, as expected from the crystal structure, and that this broken inversion symmetry combined with the mirror symmetry along the $b-c$ plane, imparts a complex spin texture that is responsible for the observed anisotropy in $H^\parallel_\textrm{c2}$. These results are in agreement with our earlier reports on monolayer MoTe$_2$ where, under the assumption of broken inversion symmetry, the tilted spin texture gives rise to a two-fold rotational symmetry for $H_\textrm{c2}$~\cite{rhodes2021enhanced}. From the calculated spin texture in Fig. \ref{F1}, we can average across the Fermi surface for $S_x$, $S_y$, and $S_z$ components and find that the average value of $S_x$ is greater than that of $S_y$. Like in the case of type-I Ising SOC, the tilting given by the finite $S_x$ value for individual electron pockets will impart a resistance for spins to align away from the $x$-direction. As a result, $H^\parallel_\textrm{c2}$ will maximize perpendicular to the $a$-axis (i.e, $H\parallel b$), in agreement with our observations. The origins of this difference in spin textures between $S_x$ and $S_y$ can be traced back to the breaking of the out-of-plane mirror symmetry given by the bilayer crystal structure.~\cite{cui2019transport}\\
\indent The lack of change to the rotational symmetry of the superconducting gap for in-plane magnetic fields through the ferroelectric transition pathway is not surprising. The moderate carrier densities of bilayer MoTe$_2$ significantly screen out-of-plane electric fields and, as a result, minimize any additional contribution to Rashba spin-orbit coupling. To see this more clearly, in Fig. \ref{F4}a we compare the DFT-calculated electronic band structure for applied out-of-plane electric fields (-0.5, 0.0, and 0.5 V/nm). While there is a minor shift in the energy of the bands (up to 2 meV), there is no appreciable change in the spin-splitting between spin-split conduction and valence bands near $E_F$. As expected, when re-examining the Fermi surfaces and concomitant spin textures as a function of applied out-of-plane electric fields, there is no change in the symmetry of the spin texture, nor is there any discernible change in the size of the Fermi pockets and maximum values of $S_x$, $S_y$, and $S_z$ for the range of out-of-plane electric fields that we have calculated. One caveat to this interpretation may be the existence of a paraelectric phase along the ferroelectric transition pathway, which has been suggested for bilayer WTe$_2$~\cite{yang2018origin,liu2019vertical} (isostructural to bilayer MoTe$_2$). Assuming this paraelectric structure, we again calculate the electronic band structure and spin texture (see Supplemental Material). While we find little change in the electronic band structure from that of the polar one, the magnitudes of $S_x$ and $S_y$ significantly decrease for the electron pockets, while remaining similar for the hole pockets. This suggests that if any intermediate metastable paraelectric phase existed along the transition pathway, we should observe the emergence of an isotropic superconducting gap. However, as pointed out in Fig. \ref{F3}d-g, we observe no collapse of the anisotropy for any value of displacement field. This suggests that domains of internal polarization may be remaining as ``up" or ``down" until sufficient displacement field of opposite sign is reached and the entire domain flips at once, instead of passing through an intermediate non-polar structure. However, more work is necessary to determine the true nature of the ferroelectric transition pathway. 
%%%%%%%%%%%%%%%%%%%%%%%%%%%%%  Summary section  %%%%%%%%%%%%%%%%%%%%%%%%%%%%%%%%%%%%%%

\indent In summary, we observe a robust two-fold anisotropic superconducting behavior in bilayer $T_d$-MoTe$_2$, where the upper critical field maximizes and minimizes with respect to the in-plane crystal symmetry. This two-fold rotational symmetry of the superconducting gap persists for all values of temperature, parallel magnetic field, applied displacement field, carrier doping, and internal polarization switching within the superconducting regime. The lack of a change in the rotational symmetry agrees with the calculated electronic band structures under electric field for the polar structure. While our evidence strongly suggests that tilted Ising SOC is responsible for the two-fold rotational symmetry, we note that it is not well understood whether other competing mechanisms can coexist to enhance the upper critical along the same axes (\textit{e.g.}, SOPC combined with tilted Ising SOC). Future work on other centro- and noncentrosymmetric layer numbers of $T_d$-MoTe$_2$ may elucidate more on these effects. We also note that in our earlier work we observed large changes in $H_\textrm{c2}$ and $T_\textrm{c}$ throughout the ferroelectric hysteresis loop, which we surmised were the result of an increase in Fermi surface nesting as the hole pocket massively increases in size while the bilayer transitions through the loop~\cite{jindal2023coupled}. While measurements of the carrier density indicate that this interpretation may still be correct, our calculations presented in this Letter show that electric fields alone cannot cause such a change to the size of the hole pockets. 

%%%%%%%%%%%%%%%%%%%%%%%%%%%%%%%%%%%%%%%%%%%%%%%%%%%%%%%%%%%%%%%
\begin{acknowledgments}
We thank Rafael M. Fernandes and Alex Levchenko for fruitful discussions. This work was supported by the Department of Energy Office of Basic Energy Sciences (DE-SC0023866), including theory (X.Q. and A.S.), measurements (D.R. and Z.L.), and analysis (Z.L., D.R., A.S., and X.Q.) and by the NSF MRSEC program through Columbia University in the Center for Precision-Assembled Quantum Materials (DMR-2011738) (A.J., C.R.D., and A.P.N.). Additional measurement support (W.Z. and L.B.) were supported by the NSF Division of Materials Research (DMR-2219003). Growth of $h$-BN crystals was supported by the Elemental Strategy Initiative conducted by the MEXT, Japan (Grant Number JPMXP0112101001), JSPS KAKENHI (Grant Numbers 19H05790, 20H00354 and 21H05233) and A3 Foresight by JSPS (T.T. and K.W.). Growth of MoTe$_2$ single crystals was supported by the Wisconsin Alumni Research Foundation (Y.H.). A portion of this work was performed at the National High Magnetic Field Laboratory, which is supported by the NSF Cooperative Agreement No. DMR-2128556 and the State of Florida. Portions of this research were conducted with the advanced computing resources provided by Texas A\&M High Performance Research Computing.
\end{acknowledgments}

\bibliographystyle{apsrev4-2}
\bibliography{Main} % Produces the bibliography via BibTeX.

\begin{thebibliography}{34}%
\makeatletter
\providecommand \@ifxundefined [1]{%
 \@ifx{#1\undefined}
}%
\providecommand \@ifnum [1]{%
 \ifnum #1\expandafter \@firstoftwo
 \else \expandafter \@secondoftwo
 \fi
}%
\providecommand \@ifx [1]{%
 \ifx #1\expandafter \@firstoftwo
 \else \expandafter \@secondoftwo
 \fi
}%
\providecommand \natexlab [1]{#1}%
\providecommand \enquote  [1]{``#1''}%
\providecommand \bibnamefont  [1]{#1}%
\providecommand \bibfnamefont [1]{#1}%
\providecommand \citenamefont [1]{#1}%
\providecommand \href@noop [0]{\@secondoftwo}%
\providecommand \href [0]{\begingroup \@sanitize@url \@href}%
\providecommand \@href[1]{\@@startlink{#1}\@@href}%
\providecommand \@@href[1]{\endgroup#1\@@endlink}%
\providecommand \@sanitize@url [0]{\catcode `\\12\catcode `\$12\catcode `\&12\catcode `\#12\catcode `\^12\catcode `\_12\catcode `\%12\relax}%
\providecommand \@@startlink[1]{}%
\providecommand \@@endlink[0]{}%
\providecommand \url  [0]{\begingroup\@sanitize@url \@url }%
\providecommand \@url [1]{\endgroup\@href {#1}{\urlprefix }}%
\providecommand \urlprefix  [0]{URL }%
\providecommand \Eprint [0]{\href }%
\providecommand \doibase [0]{https://doi.org/}%
\providecommand \selectlanguage [0]{\@gobble}%
\providecommand \bibinfo  [0]{\@secondoftwo}%
\providecommand \bibfield  [0]{\@secondoftwo}%
\providecommand \translation [1]{[#1]}%
\providecommand \BibitemOpen [0]{}%
\providecommand \bibitemStop [0]{}%
\providecommand \bibitemNoStop [0]{.\EOS\space}%
\providecommand \EOS [0]{\spacefactor3000\relax}%
\providecommand \BibitemShut  [1]{\csname bibitem#1\endcsname}%
\let\auto@bib@innerbib\@empty
%</preamble>
\bibitem [{\citenamefont {Lu}\ \emph {et~al.}(2015)\citenamefont {Lu}, \citenamefont {Zheliuk}, \citenamefont {Leermakers}, \citenamefont {Yuan}, \citenamefont {Zeitler}, \citenamefont {Law},\ and\ \citenamefont {Ye}}]{lu2015evidence}%
  \BibitemOpen
  \bibfield  {author} {\bibinfo {author} {\bibfnamefont {J.}~\bibnamefont {Lu}}, \bibinfo {author} {\bibfnamefont {O.}~\bibnamefont {Zheliuk}}, \bibinfo {author} {\bibfnamefont {I.}~\bibnamefont {Leermakers}}, \bibinfo {author} {\bibfnamefont {N.~F.}\ \bibnamefont {Yuan}}, \bibinfo {author} {\bibfnamefont {U.}~\bibnamefont {Zeitler}}, \bibinfo {author} {\bibfnamefont {K.~T.}\ \bibnamefont {Law}},\ and\ \bibinfo {author} {\bibfnamefont {J.}~\bibnamefont {Ye}},\ }\href@noop {} {\bibfield  {journal} {\bibinfo  {journal} {Science}\ }\textbf {\bibinfo {volume} {350}},\ \bibinfo {pages} {1353} (\bibinfo {year} {2015})}\BibitemShut {NoStop}%
\bibitem [{\citenamefont {Xi}\ \emph {et~al.}(2016)\citenamefont {Xi}, \citenamefont {Wang}, \citenamefont {Zhao}, \citenamefont {Park}, \citenamefont {Law}, \citenamefont {Berger}, \citenamefont {Forr{\'o}}, \citenamefont {Shan},\ and\ \citenamefont {Mak}}]{xi2016ising}%
  \BibitemOpen
  \bibfield  {author} {\bibinfo {author} {\bibfnamefont {X.}~\bibnamefont {Xi}}, \bibinfo {author} {\bibfnamefont {Z.}~\bibnamefont {Wang}}, \bibinfo {author} {\bibfnamefont {W.}~\bibnamefont {Zhao}}, \bibinfo {author} {\bibfnamefont {J.-H.}\ \bibnamefont {Park}}, \bibinfo {author} {\bibfnamefont {K.~T.}\ \bibnamefont {Law}}, \bibinfo {author} {\bibfnamefont {H.}~\bibnamefont {Berger}}, \bibinfo {author} {\bibfnamefont {L.}~\bibnamefont {Forr{\'o}}}, \bibinfo {author} {\bibfnamefont {J.}~\bibnamefont {Shan}},\ and\ \bibinfo {author} {\bibfnamefont {K.~F.}\ \bibnamefont {Mak}},\ }\href@noop {} {\bibfield  {journal} {\bibinfo  {journal} {Nature Physics}\ }\textbf {\bibinfo {volume} {12}},\ \bibinfo {pages} {139} (\bibinfo {year} {2016})}\BibitemShut {NoStop}%
\bibitem [{\citenamefont {De~la Barrera}\ \emph {et~al.}(2018)\citenamefont {De~la Barrera}, \citenamefont {Sinko}, \citenamefont {Gopalan}, \citenamefont {Sivadas}, \citenamefont {Seyler}, \citenamefont {Watanabe}, \citenamefont {Taniguchi}, \citenamefont {Tsen}, \citenamefont {Xu}, \citenamefont {Xiao},\ and\ \citenamefont {Hunt}}]{de2018tuning}%
  \BibitemOpen
  \bibfield  {author} {\bibinfo {author} {\bibfnamefont {S.~C.}\ \bibnamefont {De~la Barrera}}, \bibinfo {author} {\bibfnamefont {M.~R.}\ \bibnamefont {Sinko}}, \bibinfo {author} {\bibfnamefont {D.~P.}\ \bibnamefont {Gopalan}}, \bibinfo {author} {\bibfnamefont {N.}~\bibnamefont {Sivadas}}, \bibinfo {author} {\bibfnamefont {K.~L.}\ \bibnamefont {Seyler}}, \bibinfo {author} {\bibfnamefont {K.}~\bibnamefont {Watanabe}}, \bibinfo {author} {\bibfnamefont {T.}~\bibnamefont {Taniguchi}}, \bibinfo {author} {\bibfnamefont {A.~W.}\ \bibnamefont {Tsen}}, \bibinfo {author} {\bibfnamefont {X.}~\bibnamefont {Xu}}, \bibinfo {author} {\bibfnamefont {D.}~\bibnamefont {Xiao}},\ and\ \bibinfo {author} {\bibfnamefont {B.~M.}\ \bibnamefont {Hunt}},\ }\href@noop {} {\bibfield  {journal} {\bibinfo  {journal} {Nature Communications}\ }\textbf {\bibinfo {volume} {9}},\ \bibinfo {pages} {1427} (\bibinfo {year} {2018})}\BibitemShut {NoStop}%
\bibitem [{\citenamefont {Wang}\ \emph {et~al.}(2019)\citenamefont {Wang}, \citenamefont {Lian}, \citenamefont {Guo}, \citenamefont {Mao}, \citenamefont {Zhang}, \citenamefont {Zhang}, \citenamefont {Gu}, \citenamefont {Xu},\ and\ \citenamefont {Duan}}]{wang2019type}%
  \BibitemOpen
  \bibfield  {author} {\bibinfo {author} {\bibfnamefont {C.}~\bibnamefont {Wang}}, \bibinfo {author} {\bibfnamefont {B.}~\bibnamefont {Lian}}, \bibinfo {author} {\bibfnamefont {X.}~\bibnamefont {Guo}}, \bibinfo {author} {\bibfnamefont {J.}~\bibnamefont {Mao}}, \bibinfo {author} {\bibfnamefont {Z.}~\bibnamefont {Zhang}}, \bibinfo {author} {\bibfnamefont {D.}~\bibnamefont {Zhang}}, \bibinfo {author} {\bibfnamefont {B.-L.}\ \bibnamefont {Gu}}, \bibinfo {author} {\bibfnamefont {Y.}~\bibnamefont {Xu}},\ and\ \bibinfo {author} {\bibfnamefont {W.}~\bibnamefont {Duan}},\ }\href@noop {} {\bibfield  {journal} {\bibinfo  {journal} {Physical Review Letters}\ }\textbf {\bibinfo {volume} {123}},\ \bibinfo {pages} {126402} (\bibinfo {year} {2019})}\BibitemShut {NoStop}%
\bibitem [{\citenamefont {Falson}\ \emph {et~al.}(2020)\citenamefont {Falson}, \citenamefont {Xu}, \citenamefont {Liao}, \citenamefont {Zang}, \citenamefont {Zhu}, \citenamefont {Wang}, \citenamefont {Zhang}, \citenamefont {Liu}, \citenamefont {Duan}, \citenamefont {He}, \citenamefont {Liu}, \citenamefont {Smet}, \citenamefont {Zhang},\ and\ \citenamefont {Xue}}]{falson2020type}%
  \BibitemOpen
  \bibfield  {author} {\bibinfo {author} {\bibfnamefont {J.}~\bibnamefont {Falson}}, \bibinfo {author} {\bibfnamefont {Y.}~\bibnamefont {Xu}}, \bibinfo {author} {\bibfnamefont {M.}~\bibnamefont {Liao}}, \bibinfo {author} {\bibfnamefont {Y.}~\bibnamefont {Zang}}, \bibinfo {author} {\bibfnamefont {K.}~\bibnamefont {Zhu}}, \bibinfo {author} {\bibfnamefont {C.}~\bibnamefont {Wang}}, \bibinfo {author} {\bibfnamefont {Z.}~\bibnamefont {Zhang}}, \bibinfo {author} {\bibfnamefont {H.}~\bibnamefont {Liu}}, \bibinfo {author} {\bibfnamefont {W.}~\bibnamefont {Duan}}, \bibinfo {author} {\bibfnamefont {K.}~\bibnamefont {He}}, \bibinfo {author} {\bibfnamefont {H.}~\bibnamefont {Liu}}, \bibinfo {author} {\bibfnamefont {J.~H.}\ \bibnamefont {Smet}}, \bibinfo {author} {\bibfnamefont {D.}~\bibnamefont {Zhang}},\ and\ \bibinfo {author} {\bibfnamefont {Q.-K.}\ \bibnamefont {Xue}},\ }\href@noop {} {\bibfield  {journal} {\bibinfo  {journal} {Science}\ }\textbf {\bibinfo {volume} {367}},\ \bibinfo {pages} {1454} (\bibinfo {year}
  {2020})}\BibitemShut {NoStop}%
\bibitem [{\citenamefont {Liu}\ \emph {et~al.}(2020)\citenamefont {Liu}, \citenamefont {Xu}, \citenamefont {Sun}, \citenamefont {Liu}, \citenamefont {Liu}, \citenamefont {Wang}, \citenamefont {Zhang}, \citenamefont {Gu}, \citenamefont {Tang}, \citenamefont {Ding}, \citenamefont {Liu}, \citenamefont {Yao}, \citenamefont {Lin}, \citenamefont {Wang}, \citenamefont {Xue},\ and\ \citenamefont {Wang}}]{liu2020type}%
  \BibitemOpen
  \bibfield  {author} {\bibinfo {author} {\bibfnamefont {Y.}~\bibnamefont {Liu}}, \bibinfo {author} {\bibfnamefont {Y.}~\bibnamefont {Xu}}, \bibinfo {author} {\bibfnamefont {J.}~\bibnamefont {Sun}}, \bibinfo {author} {\bibfnamefont {C.}~\bibnamefont {Liu}}, \bibinfo {author} {\bibfnamefont {Y.}~\bibnamefont {Liu}}, \bibinfo {author} {\bibfnamefont {C.}~\bibnamefont {Wang}}, \bibinfo {author} {\bibfnamefont {Z.}~\bibnamefont {Zhang}}, \bibinfo {author} {\bibfnamefont {K.}~\bibnamefont {Gu}}, \bibinfo {author} {\bibfnamefont {Y.}~\bibnamefont {Tang}}, \bibinfo {author} {\bibfnamefont {C.}~\bibnamefont {Ding}}, \bibinfo {author} {\bibfnamefont {H.}~\bibnamefont {Liu}}, \bibinfo {author} {\bibfnamefont {H.}~\bibnamefont {Yao}}, \bibinfo {author} {\bibfnamefont {X.}~\bibnamefont {Lin}}, \bibinfo {author} {\bibfnamefont {L.}~\bibnamefont {Wang}}, \bibinfo {author} {\bibfnamefont {Q.-K.}\ \bibnamefont {Xue}},\ and\ \bibinfo {author} {\bibfnamefont {J.}~\bibnamefont {Wang}},\ }\href@noop {} {\bibfield  {journal}
  {\bibinfo  {journal} {Nano Letters}\ }\textbf {\bibinfo {volume} {20}},\ \bibinfo {pages} {5728} (\bibinfo {year} {2020})}\BibitemShut {NoStop}%
\bibitem [{\citenamefont {Cui}\ \emph {et~al.}(2019)\citenamefont {Cui}, \citenamefont {Li}, \citenamefont {Zhou}, \citenamefont {He}, \citenamefont {Huang}, \citenamefont {Yi}, \citenamefont {Fan}, \citenamefont {Ji}, \citenamefont {Jing}, \citenamefont {Qu}, \citenamefont {Cheng}, \citenamefont {Yang}, \citenamefont {Lu}, \citenamefont {Suenaga}, \citenamefont {Liu}, \citenamefont {Law}, \citenamefont {Lin}, \citenamefont {Liu},\ and\ \citenamefont {Liu}}]{cui2019transport}%
  \BibitemOpen
  \bibfield  {author} {\bibinfo {author} {\bibfnamefont {J.}~\bibnamefont {Cui}}, \bibinfo {author} {\bibfnamefont {P.}~\bibnamefont {Li}}, \bibinfo {author} {\bibfnamefont {J.}~\bibnamefont {Zhou}}, \bibinfo {author} {\bibfnamefont {W.-Y.}\ \bibnamefont {He}}, \bibinfo {author} {\bibfnamefont {X.}~\bibnamefont {Huang}}, \bibinfo {author} {\bibfnamefont {J.}~\bibnamefont {Yi}}, \bibinfo {author} {\bibfnamefont {J.}~\bibnamefont {Fan}}, \bibinfo {author} {\bibfnamefont {Z.}~\bibnamefont {Ji}}, \bibinfo {author} {\bibfnamefont {X.}~\bibnamefont {Jing}}, \bibinfo {author} {\bibfnamefont {F.}~\bibnamefont {Qu}}, \bibinfo {author} {\bibfnamefont {Z.~G.}\ \bibnamefont {Cheng}}, \bibinfo {author} {\bibfnamefont {C.}~\bibnamefont {Yang}}, \bibinfo {author} {\bibfnamefont {L.}~\bibnamefont {Lu}}, \bibinfo {author} {\bibfnamefont {K.}~\bibnamefont {Suenaga}}, \bibinfo {author} {\bibfnamefont {J.}~\bibnamefont {Liu}}, \bibinfo {author} {\bibfnamefont {K.~T.}\ \bibnamefont {Law}}, \bibinfo {author} {\bibfnamefont
  {J.}~\bibnamefont {Lin}}, \bibinfo {author} {\bibfnamefont {Z.}~\bibnamefont {Liu}},\ and\ \bibinfo {author} {\bibfnamefont {G.}~\bibnamefont {Liu}},\ }\href@noop {} {\bibfield  {journal} {\bibinfo  {journal} {Nature Communications}\ }\textbf {\bibinfo {volume} {10}},\ \bibinfo {pages} {2044} (\bibinfo {year} {2019})}\BibitemShut {NoStop}%
\bibitem [{\citenamefont {Rhodes}\ \emph {et~al.}(2021)\citenamefont {Rhodes}, \citenamefont {Jindal}, \citenamefont {Yuan}, \citenamefont {Jung}, \citenamefont {Antony}, \citenamefont {Wang}, \citenamefont {Kim}, \citenamefont {Chiu}, \citenamefont {Taniguchi}, \citenamefont {Watanabe}, \citenamefont {Barmak}, \citenamefont {Balicas}, \citenamefont {Dean}, \citenamefont {Qian}, \citenamefont {Fu}, \citenamefont {Pasupathy},\ and\ \citenamefont {Hone}}]{rhodes2021enhanced}%
  \BibitemOpen
  \bibfield  {author} {\bibinfo {author} {\bibfnamefont {D.~A.}\ \bibnamefont {Rhodes}}, \bibinfo {author} {\bibfnamefont {A.}~\bibnamefont {Jindal}}, \bibinfo {author} {\bibfnamefont {N.~F.}\ \bibnamefont {Yuan}}, \bibinfo {author} {\bibfnamefont {Y.}~\bibnamefont {Jung}}, \bibinfo {author} {\bibfnamefont {A.}~\bibnamefont {Antony}}, \bibinfo {author} {\bibfnamefont {H.}~\bibnamefont {Wang}}, \bibinfo {author} {\bibfnamefont {B.}~\bibnamefont {Kim}}, \bibinfo {author} {\bibfnamefont {Y.-c.}\ \bibnamefont {Chiu}}, \bibinfo {author} {\bibfnamefont {T.}~\bibnamefont {Taniguchi}}, \bibinfo {author} {\bibfnamefont {K.}~\bibnamefont {Watanabe}}, \bibinfo {author} {\bibfnamefont {K.}~\bibnamefont {Barmak}}, \bibinfo {author} {\bibfnamefont {L.}~\bibnamefont {Balicas}}, \bibinfo {author} {\bibfnamefont {C.~R.}\ \bibnamefont {Dean}}, \bibinfo {author} {\bibfnamefont {X.}~\bibnamefont {Qian}}, \bibinfo {author} {\bibfnamefont {L.}~\bibnamefont {Fu}}, \bibinfo {author} {\bibfnamefont {A.~N.}\ \bibnamefont
  {Pasupathy}},\ and\ \bibinfo {author} {\bibfnamefont {J.}~\bibnamefont {Hone}},\ }\href@noop {} {\bibfield  {journal} {\bibinfo  {journal} {Nano Letters}\ }\textbf {\bibinfo {volume} {21}},\ \bibinfo {pages} {2505} (\bibinfo {year} {2021})}\BibitemShut {NoStop}%
\bibitem [{\citenamefont {Yoshizawa}\ \emph {et~al.}(2021)\citenamefont {Yoshizawa}, \citenamefont {Kobayashi}, \citenamefont {Nakata}, \citenamefont {Yaji}, \citenamefont {Yokota}, \citenamefont {Komori}, \citenamefont {Shin}, \citenamefont {Sakamoto},\ and\ \citenamefont {Uchihashi}}]{yoshizawa2021atomic}%
  \BibitemOpen
  \bibfield  {author} {\bibinfo {author} {\bibfnamefont {S.}~\bibnamefont {Yoshizawa}}, \bibinfo {author} {\bibfnamefont {T.}~\bibnamefont {Kobayashi}}, \bibinfo {author} {\bibfnamefont {Y.}~\bibnamefont {Nakata}}, \bibinfo {author} {\bibfnamefont {K.}~\bibnamefont {Yaji}}, \bibinfo {author} {\bibfnamefont {K.}~\bibnamefont {Yokota}}, \bibinfo {author} {\bibfnamefont {F.}~\bibnamefont {Komori}}, \bibinfo {author} {\bibfnamefont {S.}~\bibnamefont {Shin}}, \bibinfo {author} {\bibfnamefont {K.}~\bibnamefont {Sakamoto}},\ and\ \bibinfo {author} {\bibfnamefont {T.}~\bibnamefont {Uchihashi}},\ }\href@noop {} {\bibfield  {journal} {\bibinfo  {journal} {Nature Communications}\ }\textbf {\bibinfo {volume} {12}},\ \bibinfo {pages} {1462} (\bibinfo {year} {2021})}\BibitemShut {NoStop}%
\bibitem [{\citenamefont {Xie}\ \emph {et~al.}(2020)\citenamefont {Xie}, \citenamefont {Zhou},\ and\ \citenamefont {Law}}]{xie2020spin}%
  \BibitemOpen
  \bibfield  {author} {\bibinfo {author} {\bibfnamefont {Y.-M.}\ \bibnamefont {Xie}}, \bibinfo {author} {\bibfnamefont {B.~T.}\ \bibnamefont {Zhou}},\ and\ \bibinfo {author} {\bibfnamefont {K.~T.}\ \bibnamefont {Law}},\ }\href@noop {} {\bibfield  {journal} {\bibinfo  {journal} {Physical Review Letters}\ }\textbf {\bibinfo {volume} {125}},\ \bibinfo {pages} {107001} (\bibinfo {year} {2020})}\BibitemShut {NoStop}%
\bibitem [{\citenamefont {Zhang}\ \emph {et~al.}(2023)\citenamefont {Zhang}, \citenamefont {Xie}, \citenamefont {Fang}, \citenamefont {Zhang}, \citenamefont {Xu}, \citenamefont {Zou}, \citenamefont {Leng}, \citenamefont {Gao}, \citenamefont {Zhang}, \citenamefont {Ai}, \citenamefont {Zhang}, \citenamefont {Jia}, \citenamefont {Liu}, \citenamefont {Yan}, \citenamefont {Zhao}, \citenamefont {Haigh}, \citenamefont {Kou}, \citenamefont {Yang}, \citenamefont {Huang}, \citenamefont {Law}, \citenamefont {Xiu},\ and\ \citenamefont {Dong}}]{zhang2023spin}%
  \BibitemOpen
  \bibfield  {author} {\bibinfo {author} {\bibfnamefont {E.}~\bibnamefont {Zhang}}, \bibinfo {author} {\bibfnamefont {Y.-M.}\ \bibnamefont {Xie}}, \bibinfo {author} {\bibfnamefont {Y.}~\bibnamefont {Fang}}, \bibinfo {author} {\bibfnamefont {J.}~\bibnamefont {Zhang}}, \bibinfo {author} {\bibfnamefont {X.}~\bibnamefont {Xu}}, \bibinfo {author} {\bibfnamefont {Y.-C.}\ \bibnamefont {Zou}}, \bibinfo {author} {\bibfnamefont {P.}~\bibnamefont {Leng}}, \bibinfo {author} {\bibfnamefont {X.-J.}\ \bibnamefont {Gao}}, \bibinfo {author} {\bibfnamefont {Y.}~\bibnamefont {Zhang}}, \bibinfo {author} {\bibfnamefont {L.}~\bibnamefont {Ai}}, \bibinfo {author} {\bibfnamefont {Y.}~\bibnamefont {Zhang}}, \bibinfo {author} {\bibfnamefont {Z.}~\bibnamefont {Jia}}, \bibinfo {author} {\bibfnamefont {S.}~\bibnamefont {Liu}}, \bibinfo {author} {\bibfnamefont {J.}~\bibnamefont {Yan}}, \bibinfo {author} {\bibfnamefont {W.}~\bibnamefont {Zhao}}, \bibinfo {author} {\bibfnamefont {S.~J.}\ \bibnamefont {Haigh}}, \bibinfo {author}
  {\bibfnamefont {X.}~\bibnamefont {Kou}}, \bibinfo {author} {\bibfnamefont {J.}~\bibnamefont {Yang}}, \bibinfo {author} {\bibfnamefont {F.}~\bibnamefont {Huang}}, \bibinfo {author} {\bibfnamefont {K.~T.}\ \bibnamefont {Law}}, \bibinfo {author} {\bibfnamefont {F.}~\bibnamefont {Xiu}},\ and\ \bibinfo {author} {\bibfnamefont {S.}~\bibnamefont {Dong}},\ }\href@noop {} {\bibfield  {journal} {\bibinfo  {journal} {Nature Physics}\ }\textbf {\bibinfo {volume} {19}},\ \bibinfo {pages} {106} (\bibinfo {year} {2023})}\BibitemShut {NoStop}%
\bibitem [{\citenamefont {Qian}\ \emph {et~al.}(2014)\citenamefont {Qian}, \citenamefont {Liu}, \citenamefont {Fu},\ and\ \citenamefont {Li}}]{qian2014quantum}%
  \BibitemOpen
  \bibfield  {author} {\bibinfo {author} {\bibfnamefont {X.}~\bibnamefont {Qian}}, \bibinfo {author} {\bibfnamefont {J.}~\bibnamefont {Liu}}, \bibinfo {author} {\bibfnamefont {L.}~\bibnamefont {Fu}},\ and\ \bibinfo {author} {\bibfnamefont {J.}~\bibnamefont {Li}},\ }\href@noop {} {\bibfield  {journal} {\bibinfo  {journal} {Science}\ }\textbf {\bibinfo {volume} {346}},\ \bibinfo {pages} {1344} (\bibinfo {year} {2014})}\BibitemShut {NoStop}%
\bibitem [{\citenamefont {Tang}\ \emph {et~al.}(2017)\citenamefont {Tang}, \citenamefont {Zhang}, \citenamefont {Wong}, \citenamefont {Pedramrazi}, \citenamefont {Tsai}, \citenamefont {Jia}, \citenamefont {Moritz}, \citenamefont {Claassen}, \citenamefont {Ryu}, \citenamefont {Kahn}, \citenamefont {Jiang}, \citenamefont {Yan}, \citenamefont {Hashimoto}, \citenamefont {Lu}, \citenamefont {Moore}, \citenamefont {Chan-Cuk}, \citenamefont {Hwang}, \citenamefont {Hussain}, \citenamefont {Chen}, \citenamefont {Ugeda}, \citenamefont {Liu}, \citenamefont {Xie}, \citenamefont {Devereaux}, \citenamefont {Crommie}, \citenamefont {Mo},\ and\ \citenamefont {Shen}}]{tang2017quantum}%
  \BibitemOpen
  \bibfield  {author} {\bibinfo {author} {\bibfnamefont {S.}~\bibnamefont {Tang}}, \bibinfo {author} {\bibfnamefont {C.}~\bibnamefont {Zhang}}, \bibinfo {author} {\bibfnamefont {D.}~\bibnamefont {Wong}}, \bibinfo {author} {\bibfnamefont {Z.}~\bibnamefont {Pedramrazi}}, \bibinfo {author} {\bibfnamefont {H.-Z.}\ \bibnamefont {Tsai}}, \bibinfo {author} {\bibfnamefont {C.}~\bibnamefont {Jia}}, \bibinfo {author} {\bibfnamefont {B.}~\bibnamefont {Moritz}}, \bibinfo {author} {\bibfnamefont {M.}~\bibnamefont {Claassen}}, \bibinfo {author} {\bibfnamefont {H.}~\bibnamefont {Ryu}}, \bibinfo {author} {\bibfnamefont {S.}~\bibnamefont {Kahn}}, \bibinfo {author} {\bibfnamefont {J.}~\bibnamefont {Jiang}}, \bibinfo {author} {\bibfnamefont {H.}~\bibnamefont {Yan}}, \bibinfo {author} {\bibfnamefont {M.}~\bibnamefont {Hashimoto}}, \bibinfo {author} {\bibfnamefont {D.}~\bibnamefont {Lu}}, \bibinfo {author} {\bibfnamefont {R.~G.}\ \bibnamefont {Moore}}, \bibinfo {author} {\bibfnamefont {H.}~\bibnamefont {Chan-Cuk}}, \bibinfo
  {author} {\bibfnamefont {C.}~\bibnamefont {Hwang}}, \bibinfo {author} {\bibfnamefont {Z.}~\bibnamefont {Hussain}}, \bibinfo {author} {\bibfnamefont {Y.}~\bibnamefont {Chen}}, \bibinfo {author} {\bibfnamefont {M.~M.}\ \bibnamefont {Ugeda}}, \bibinfo {author} {\bibfnamefont {Z.}~\bibnamefont {Liu}}, \bibinfo {author} {\bibfnamefont {X.}~\bibnamefont {Xie}}, \bibinfo {author} {\bibfnamefont {T.~P.}\ \bibnamefont {Devereaux}}, \bibinfo {author} {\bibfnamefont {M.~F.}\ \bibnamefont {Crommie}}, \bibinfo {author} {\bibfnamefont {S.~K.}\ \bibnamefont {Mo}},\ and\ \bibinfo {author} {\bibfnamefont {Z.~X.}\ \bibnamefont {Shen}},\ }\href@noop {} {\bibfield  {journal} {\bibinfo  {journal} {Nature Physics}\ }\textbf {\bibinfo {volume} {13}},\ \bibinfo {pages} {683} (\bibinfo {year} {2017})}\BibitemShut {NoStop}%
\bibitem [{\citenamefont {Sato}\ and\ \citenamefont {Ando}(2017)}]{sato2017topological}%
  \BibitemOpen
  \bibfield  {author} {\bibinfo {author} {\bibfnamefont {M.}~\bibnamefont {Sato}}\ and\ \bibinfo {author} {\bibfnamefont {Y.}~\bibnamefont {Ando}},\ }\href@noop {} {\bibfield  {journal} {\bibinfo  {journal} {Reports on Progress in Physics}\ }\textbf {\bibinfo {volume} {80}},\ \bibinfo {pages} {076501} (\bibinfo {year} {2017})}\BibitemShut {NoStop}%
\bibitem [{\citenamefont {Lee}\ and\ \citenamefont {Son}(2021)}]{lee2021gate}%
  \BibitemOpen
  \bibfield  {author} {\bibinfo {author} {\bibfnamefont {J.-H.}\ \bibnamefont {Lee}}\ and\ \bibinfo {author} {\bibfnamefont {Y.-W.}\ \bibnamefont {Son}},\ }\href@noop {} {\bibfield  {journal} {\bibinfo  {journal} {Physical Chemistry Chemical Physics}\ }\textbf {\bibinfo {volume} {23}},\ \bibinfo {pages} {17279} (\bibinfo {year} {2021})}\BibitemShut {NoStop}%
\bibitem [{\citenamefont {Liu}\ \emph {et~al.}(2021)\citenamefont {Liu}, \citenamefont {Chong}, \citenamefont {Sharma},\ and\ \citenamefont {Davis}}]{liu2021discovery}%
  \BibitemOpen
  \bibfield  {author} {\bibinfo {author} {\bibfnamefont {X.}~\bibnamefont {Liu}}, \bibinfo {author} {\bibfnamefont {Y.~X.}\ \bibnamefont {Chong}}, \bibinfo {author} {\bibfnamefont {R.}~\bibnamefont {Sharma}},\ and\ \bibinfo {author} {\bibfnamefont {J.~S.}\ \bibnamefont {Davis}},\ }\href@noop {} {\bibfield  {journal} {\bibinfo  {journal} {Science}\ }\textbf {\bibinfo {volume} {372}},\ \bibinfo {pages} {1447} (\bibinfo {year} {2021})}\BibitemShut {NoStop}%
\bibitem [{\citenamefont {Wei}\ \emph {et~al.}(2023)\citenamefont {Wei}, \citenamefont {Xiao}, \citenamefont {Li}, \citenamefont {Wang}, \citenamefont {Deng}, \citenamefont {Cheng}, \citenamefont {Zheng}, \citenamefont {Hao}, \citenamefont {Zhang}, \citenamefont {Ma}, \citenamefont {Xue},\ and\ \citenamefont {Song}}]{wei2023discovery}%
  \BibitemOpen
  \bibfield  {author} {\bibinfo {author} {\bibfnamefont {L.-X.}\ \bibnamefont {Wei}}, \bibinfo {author} {\bibfnamefont {P.-C.}\ \bibnamefont {Xiao}}, \bibinfo {author} {\bibfnamefont {F.}~\bibnamefont {Li}}, \bibinfo {author} {\bibfnamefont {L.}~\bibnamefont {Wang}}, \bibinfo {author} {\bibfnamefont {B.-Y.}\ \bibnamefont {Deng}}, \bibinfo {author} {\bibfnamefont {F.-J.}\ \bibnamefont {Cheng}}, \bibinfo {author} {\bibfnamefont {F.-W.}\ \bibnamefont {Zheng}}, \bibinfo {author} {\bibfnamefont {N.}~\bibnamefont {Hao}}, \bibinfo {author} {\bibfnamefont {P.}~\bibnamefont {Zhang}}, \bibinfo {author} {\bibfnamefont {X.-C.}\ \bibnamefont {Ma}}, \bibinfo {author} {\bibfnamefont {Q.-K.}\ \bibnamefont {Xue}},\ and\ \bibinfo {author} {\bibfnamefont {C.-L.}\ \bibnamefont {Song}},\ }\href@noop {} {\bibfield  {journal} {\bibinfo  {journal} {arXiv preprint arXiv:2308.11101}\ } (\bibinfo {year} {2023})}\BibitemShut {NoStop}%
\bibitem [{\citenamefont {Wan}\ \emph {et~al.}(2023)\citenamefont {Wan}, \citenamefont {Zheliuk}, \citenamefont {Yuan}, \citenamefont {Peng}, \citenamefont {Zhang}, \citenamefont {Liang}, \citenamefont {Zeitler}, \citenamefont {Wiedmann}, \citenamefont {Hussey}, \citenamefont {Palstra},\ and\ \citenamefont {Ye}}]{wan2023orbital}%
  \BibitemOpen
  \bibfield  {author} {\bibinfo {author} {\bibfnamefont {P.}~\bibnamefont {Wan}}, \bibinfo {author} {\bibfnamefont {O.}~\bibnamefont {Zheliuk}}, \bibinfo {author} {\bibfnamefont {N.~F.}\ \bibnamefont {Yuan}}, \bibinfo {author} {\bibfnamefont {X.}~\bibnamefont {Peng}}, \bibinfo {author} {\bibfnamefont {L.}~\bibnamefont {Zhang}}, \bibinfo {author} {\bibfnamefont {M.}~\bibnamefont {Liang}}, \bibinfo {author} {\bibfnamefont {U.}~\bibnamefont {Zeitler}}, \bibinfo {author} {\bibfnamefont {S.}~\bibnamefont {Wiedmann}}, \bibinfo {author} {\bibfnamefont {N.~E.}\ \bibnamefont {Hussey}}, \bibinfo {author} {\bibfnamefont {T.~T.}\ \bibnamefont {Palstra}},\ and\ \bibinfo {author} {\bibfnamefont {J.}~\bibnamefont {Ye}},\ }\href@noop {} {\bibfield  {journal} {\bibinfo  {journal} {Nature}\ }\textbf {\bibinfo {volume} {619}},\ \bibinfo {pages} {46} (\bibinfo {year} {2023})}\BibitemShut {NoStop}%
\bibitem [{\citenamefont {Jindal}\ \emph {et~al.}(2023)\citenamefont {Jindal}, \citenamefont {Saha}, \citenamefont {Li}, \citenamefont {Taniguchi}, \citenamefont {Watanabe}, \citenamefont {Hone}, \citenamefont {Birol}, \citenamefont {Fernandes}, \citenamefont {Dean}, \citenamefont {Pasupathy},\ and\ \citenamefont {A}}]{jindal2023coupled}%
  \BibitemOpen
  \bibfield  {author} {\bibinfo {author} {\bibfnamefont {A.}~\bibnamefont {Jindal}}, \bibinfo {author} {\bibfnamefont {A.}~\bibnamefont {Saha}}, \bibinfo {author} {\bibfnamefont {Z.}~\bibnamefont {Li}}, \bibinfo {author} {\bibfnamefont {T.}~\bibnamefont {Taniguchi}}, \bibinfo {author} {\bibfnamefont {K.}~\bibnamefont {Watanabe}}, \bibinfo {author} {\bibfnamefont {J.~C.}\ \bibnamefont {Hone}}, \bibinfo {author} {\bibfnamefont {T.}~\bibnamefont {Birol}}, \bibinfo {author} {\bibfnamefont {R.~M.}\ \bibnamefont {Fernandes}}, \bibinfo {author} {\bibfnamefont {C.~R.}\ \bibnamefont {Dean}}, \bibinfo {author} {\bibfnamefont {A.~N.}\ \bibnamefont {Pasupathy}},\ and\ \bibinfo {author} {\bibfnamefont {R.}~\bibnamefont {A}},\ }\href@noop {} {\bibfield  {journal} {\bibinfo  {journal} {Nature}\ }\textbf {\bibinfo {volume} {613}},\ \bibinfo {pages} {48} (\bibinfo {year} {2023})}\BibitemShut {NoStop}%
\bibitem [{\citenamefont {Chen}\ \emph {et~al.}(2016)\citenamefont {Chen}, \citenamefont {Lv}, \citenamefont {Luo}, \citenamefont {Lu}, \citenamefont {Pei}, \citenamefont {Lin}, \citenamefont {Han}, \citenamefont {Zhu}, \citenamefont {Song},\ and\ \citenamefont {Sun}}]{chen2016extremely}%
  \BibitemOpen
  \bibfield  {author} {\bibinfo {author} {\bibfnamefont {F.}~\bibnamefont {Chen}}, \bibinfo {author} {\bibfnamefont {H.}~\bibnamefont {Lv}}, \bibinfo {author} {\bibfnamefont {X.}~\bibnamefont {Luo}}, \bibinfo {author} {\bibfnamefont {W.}~\bibnamefont {Lu}}, \bibinfo {author} {\bibfnamefont {Q.}~\bibnamefont {Pei}}, \bibinfo {author} {\bibfnamefont {G.}~\bibnamefont {Lin}}, \bibinfo {author} {\bibfnamefont {Y.}~\bibnamefont {Han}}, \bibinfo {author} {\bibfnamefont {X.}~\bibnamefont {Zhu}}, \bibinfo {author} {\bibfnamefont {W.}~\bibnamefont {Song}},\ and\ \bibinfo {author} {\bibfnamefont {Y.}~\bibnamefont {Sun}},\ }\href@noop {} {\bibfield  {journal} {\bibinfo  {journal} {Physical Review B}\ }\textbf {\bibinfo {volume} {94}},\ \bibinfo {pages} {235154} (\bibinfo {year} {2016})}\BibitemShut {NoStop}%
\bibitem [{\citenamefont {Tinkham}(2004)}]{tinkham2004introduction}%
  \BibitemOpen
  \bibfield  {author} {\bibinfo {author} {\bibfnamefont {M.}~\bibnamefont {Tinkham}},\ }\href@noop {} {\emph {\bibinfo {title} {{Introduction to Superconductivity}}}}\ (\bibinfo  {publisher} {Courier Corporation},\ \bibinfo {year} {2004})\BibitemShut {NoStop}%
\bibitem [{\citenamefont {Saito}\ \emph {et~al.}(2016)\citenamefont {Saito}, \citenamefont {Nojima},\ and\ \citenamefont {Iwasa}}]{saito2016highly}%
  \BibitemOpen
  \bibfield  {author} {\bibinfo {author} {\bibfnamefont {Y.}~\bibnamefont {Saito}}, \bibinfo {author} {\bibfnamefont {T.}~\bibnamefont {Nojima}},\ and\ \bibinfo {author} {\bibfnamefont {Y.}~\bibnamefont {Iwasa}},\ }\href@noop {} {\bibfield  {journal} {\bibinfo  {journal} {Nature Reviews Materials}\ }\textbf {\bibinfo {volume} {2}},\ \bibinfo {pages} {16094} (\bibinfo {year} {2016})}\BibitemShut {NoStop}%
\bibitem [{\citenamefont {Hamill}\ \emph {et~al.}(2021)\citenamefont {Hamill}, \citenamefont {Heischmidt}, \citenamefont {Sohn}, \citenamefont {Shaffer}, \citenamefont {Tsai}, \citenamefont {Zhang}, \citenamefont {Xi}, \citenamefont {Suslov}, \citenamefont {Berger}, \citenamefont {Forr{\'o}}, \citenamefont {Burnell}, \citenamefont {Shan}, \citenamefont {Mak}, \citenamefont {Fernandes}, \citenamefont {Wang},\ and\ \citenamefont {Pribiag}}]{hamill2021two}%
  \BibitemOpen
  \bibfield  {author} {\bibinfo {author} {\bibfnamefont {A.}~\bibnamefont {Hamill}}, \bibinfo {author} {\bibfnamefont {B.}~\bibnamefont {Heischmidt}}, \bibinfo {author} {\bibfnamefont {E.}~\bibnamefont {Sohn}}, \bibinfo {author} {\bibfnamefont {D.}~\bibnamefont {Shaffer}}, \bibinfo {author} {\bibfnamefont {K.-T.}\ \bibnamefont {Tsai}}, \bibinfo {author} {\bibfnamefont {X.}~\bibnamefont {Zhang}}, \bibinfo {author} {\bibfnamefont {X.}~\bibnamefont {Xi}}, \bibinfo {author} {\bibfnamefont {A.}~\bibnamefont {Suslov}}, \bibinfo {author} {\bibfnamefont {H.}~\bibnamefont {Berger}}, \bibinfo {author} {\bibfnamefont {L.}~\bibnamefont {Forr{\'o}}}, \bibinfo {author} {\bibfnamefont {F.~J.}\ \bibnamefont {Burnell}}, \bibinfo {author} {\bibfnamefont {J.}~\bibnamefont {Shan}}, \bibinfo {author} {\bibfnamefont {K.~F.}\ \bibnamefont {Mak}}, \bibinfo {author} {\bibfnamefont {R.~M.}\ \bibnamefont {Fernandes}}, \bibinfo {author} {\bibfnamefont {K.}~\bibnamefont {Wang}},\ and\ \bibinfo {author} {\bibfnamefont {V.~S.}\
  \bibnamefont {Pribiag}},\ }\href@noop {} {\bibfield  {journal} {\bibinfo  {journal} {Nature Physics}\ }\textbf {\bibinfo {volume} {17}},\ \bibinfo {pages} {949} (\bibinfo {year} {2021})}\BibitemShut {NoStop}%
\bibitem [{\citenamefont {Beams}\ \emph {et~al.}(2016)\citenamefont {Beams}, \citenamefont {Can{\c{c}}ado}, \citenamefont {Krylyuk}, \citenamefont {Kalish}, \citenamefont {Kalanyan}, \citenamefont {Singh}, \citenamefont {Choudhary}, \citenamefont {Bruma}, \citenamefont {Vora}, \citenamefont {Tavazza}, \citenamefont {Davydov},\ and\ \citenamefont {Stranick}}]{beams2016characterization}%
  \BibitemOpen
  \bibfield  {author} {\bibinfo {author} {\bibfnamefont {R.}~\bibnamefont {Beams}}, \bibinfo {author} {\bibfnamefont {L.~G.}\ \bibnamefont {Can{\c{c}}ado}}, \bibinfo {author} {\bibfnamefont {S.}~\bibnamefont {Krylyuk}}, \bibinfo {author} {\bibfnamefont {I.}~\bibnamefont {Kalish}}, \bibinfo {author} {\bibfnamefont {B.}~\bibnamefont {Kalanyan}}, \bibinfo {author} {\bibfnamefont {A.~K.}\ \bibnamefont {Singh}}, \bibinfo {author} {\bibfnamefont {K.}~\bibnamefont {Choudhary}}, \bibinfo {author} {\bibfnamefont {A.}~\bibnamefont {Bruma}}, \bibinfo {author} {\bibfnamefont {P.~M.}\ \bibnamefont {Vora}}, \bibinfo {author} {\bibfnamefont {F.}~\bibnamefont {Tavazza}}, \bibinfo {author} {\bibfnamefont {A.~V.}\ \bibnamefont {Davydov}},\ and\ \bibinfo {author} {\bibfnamefont {S.~J.}\ \bibnamefont {Stranick}},\ }\href@noop {} {\bibfield  {journal} {\bibinfo  {journal} {ACS Nano}\ }\textbf {\bibinfo {volume} {10}},\ \bibinfo {pages} {9626} (\bibinfo {year} {2016})}\BibitemShut {NoStop}%
\bibitem [{\citenamefont {Sigrist}(2009)}]{sigrist2009introduction}%
  \BibitemOpen
  \bibfield  {author} {\bibinfo {author} {\bibfnamefont {M.}~\bibnamefont {Sigrist}},\ }\href@noop {} {\bibfield  {journal} {\bibinfo  {journal} {AIP Conference Proceedings}\ }\textbf {\bibinfo {volume} {1162}},\ \bibinfo {pages} {55} (\bibinfo {year} {2009})}\BibitemShut {NoStop}%
\bibitem [{\citenamefont {Youn}\ \emph {et~al.}(2012)\citenamefont {Youn}, \citenamefont {Fischer}, \citenamefont {Rhim}, \citenamefont {Sigrist},\ and\ \citenamefont {Agterberg}}]{youn2012role}%
  \BibitemOpen
  \bibfield  {author} {\bibinfo {author} {\bibfnamefont {S.~J.}\ \bibnamefont {Youn}}, \bibinfo {author} {\bibfnamefont {M.~H.}\ \bibnamefont {Fischer}}, \bibinfo {author} {\bibfnamefont {S.}~\bibnamefont {Rhim}}, \bibinfo {author} {\bibfnamefont {M.}~\bibnamefont {Sigrist}},\ and\ \bibinfo {author} {\bibfnamefont {D.~F.}\ \bibnamefont {Agterberg}},\ }\href@noop {} {\bibfield  {journal} {\bibinfo  {journal} {Physical Review B}\ }\textbf {\bibinfo {volume} {85}},\ \bibinfo {pages} {220505} (\bibinfo {year} {2012})}\BibitemShut {NoStop}%
\bibitem [{\citenamefont {Galitski}\ and\ \citenamefont {Larkin}(2001)}]{galitski2001disorder}%
  \BibitemOpen
  \bibfield  {author} {\bibinfo {author} {\bibfnamefont {V.}~\bibnamefont {Galitski}}\ and\ \bibinfo {author} {\bibfnamefont {A.}~\bibnamefont {Larkin}},\ }\href@noop {} {\bibfield  {journal} {\bibinfo  {journal} {Physical Review Letters}\ }\textbf {\bibinfo {volume} {87}},\ \bibinfo {pages} {087001} (\bibinfo {year} {2001})}\BibitemShut {NoStop}%
\bibitem [{\citenamefont {Rhodes}\ \emph {et~al.}(2017)\citenamefont {Rhodes}, \citenamefont {Sch{\"o}nemann}, \citenamefont {Aryal}, \citenamefont {Zhou}, \citenamefont {Zhang}, \citenamefont {Kampert}, \citenamefont {Chiu}, \citenamefont {Lai}, \citenamefont {Shimura}, \citenamefont {McCandless}, \citenamefont {Chan}, \citenamefont {Paley}, \citenamefont {J}, \citenamefont {Finke}, \citenamefont {Ruff}, \citenamefont {Das}, \citenamefont {Manousakis},\ and\ \citenamefont {Balicas}}]{rhodes2017bulk}%
  \BibitemOpen
  \bibfield  {author} {\bibinfo {author} {\bibfnamefont {D.}~\bibnamefont {Rhodes}}, \bibinfo {author} {\bibfnamefont {R.}~\bibnamefont {Sch{\"o}nemann}}, \bibinfo {author} {\bibfnamefont {N.}~\bibnamefont {Aryal}}, \bibinfo {author} {\bibfnamefont {Q.}~\bibnamefont {Zhou}}, \bibinfo {author} {\bibfnamefont {Q.}~\bibnamefont {Zhang}}, \bibinfo {author} {\bibfnamefont {E.}~\bibnamefont {Kampert}}, \bibinfo {author} {\bibfnamefont {Y.-C.}\ \bibnamefont {Chiu}}, \bibinfo {author} {\bibfnamefont {Y.}~\bibnamefont {Lai}}, \bibinfo {author} {\bibfnamefont {Y.}~\bibnamefont {Shimura}}, \bibinfo {author} {\bibfnamefont {G.}~\bibnamefont {McCandless}}, \bibinfo {author} {\bibfnamefont {J.}~\bibnamefont {Chan}}, \bibinfo {author} {\bibfnamefont {D.}~\bibnamefont {Paley}}, \bibinfo {author} {\bibfnamefont {L.}~\bibnamefont {J}}, \bibinfo {author} {\bibfnamefont {A.}~\bibnamefont {Finke}}, \bibinfo {author} {\bibfnamefont {J.}~\bibnamefont {Ruff}}, \bibinfo {author} {\bibfnamefont {S.}~\bibnamefont {Das}}, \bibinfo
  {author} {\bibfnamefont {E.}~\bibnamefont {Manousakis}},\ and\ \bibinfo {author} {\bibfnamefont {L.}~\bibnamefont {Balicas}},\ }\href@noop {} {\bibfield  {journal} {\bibinfo  {journal} {Physical Review B}\ }\textbf {\bibinfo {volume} {96}},\ \bibinfo {pages} {165134} (\bibinfo {year} {2017})}\BibitemShut {NoStop}%
\bibitem [{\citenamefont {Klemm}\ \emph {et~al.}(1975)\citenamefont {Klemm}, \citenamefont {Luther},\ and\ \citenamefont {Beasley}}]{klemm1975theory}%
  \BibitemOpen
  \bibfield  {author} {\bibinfo {author} {\bibfnamefont {R.~A.}\ \bibnamefont {Klemm}}, \bibinfo {author} {\bibfnamefont {A.}~\bibnamefont {Luther}},\ and\ \bibinfo {author} {\bibfnamefont {M.}~\bibnamefont {Beasley}},\ }\href@noop {} {\bibfield  {journal} {\bibinfo  {journal} {Physical Review B}\ }\textbf {\bibinfo {volume} {12}},\ \bibinfo {pages} {877} (\bibinfo {year} {1975})}\BibitemShut {NoStop}%
\bibitem [{\citenamefont {Cheon}\ \emph {et~al.}(2021)\citenamefont {Cheon}, \citenamefont {Lim}, \citenamefont {Kim},\ and\ \citenamefont {Cheong}}]{cheon2021structural}%
  \BibitemOpen
  \bibfield  {author} {\bibinfo {author} {\bibfnamefont {Y.}~\bibnamefont {Cheon}}, \bibinfo {author} {\bibfnamefont {S.~Y.}\ \bibnamefont {Lim}}, \bibinfo {author} {\bibfnamefont {K.}~\bibnamefont {Kim}},\ and\ \bibinfo {author} {\bibfnamefont {H.}~\bibnamefont {Cheong}},\ }\href@noop {} {\bibfield  {journal} {\bibinfo  {journal} {ACS Nano}\ }\textbf {\bibinfo {volume} {15}},\ \bibinfo {pages} {2962} (\bibinfo {year} {2021})}\BibitemShut {NoStop}%
\bibitem [{\citenamefont {Hart}\ \emph {et~al.}(2023)\citenamefont {Hart}, \citenamefont {Bhatt}, \citenamefont {Zhu}, \citenamefont {Han}, \citenamefont {Bianco}, \citenamefont {Li}, \citenamefont {Hynek}, \citenamefont {Schneeloch}, \citenamefont {Tao}, \citenamefont {Louca}, \citenamefont {Guo}, \citenamefont {Zhu}, \citenamefont {Jornada}, \citenamefont {Reed}, \citenamefont {Kourkoutis},\ and\ \citenamefont {Cha}}]{hart2023emergent}%
  \BibitemOpen
  \bibfield  {author} {\bibinfo {author} {\bibfnamefont {J.~L.}\ \bibnamefont {Hart}}, \bibinfo {author} {\bibfnamefont {L.}~\bibnamefont {Bhatt}}, \bibinfo {author} {\bibfnamefont {Y.}~\bibnamefont {Zhu}}, \bibinfo {author} {\bibfnamefont {M.-G.}\ \bibnamefont {Han}}, \bibinfo {author} {\bibfnamefont {E.}~\bibnamefont {Bianco}}, \bibinfo {author} {\bibfnamefont {S.}~\bibnamefont {Li}}, \bibinfo {author} {\bibfnamefont {D.~J.}\ \bibnamefont {Hynek}}, \bibinfo {author} {\bibfnamefont {J.~A.}\ \bibnamefont {Schneeloch}}, \bibinfo {author} {\bibfnamefont {Y.}~\bibnamefont {Tao}}, \bibinfo {author} {\bibfnamefont {D.}~\bibnamefont {Louca}}, \bibinfo {author} {\bibfnamefont {P.}~\bibnamefont {Guo}}, \bibinfo {author} {\bibfnamefont {Y.}~\bibnamefont {Zhu}}, \bibinfo {author} {\bibfnamefont {F.}~\bibnamefont {Jornada}}, \bibinfo {author} {\bibfnamefont {E.~J.}\ \bibnamefont {Reed}}, \bibinfo {author} {\bibfnamefont {L.~F.}\ \bibnamefont {Kourkoutis}},\ and\ \bibinfo {author} {\bibfnamefont {J.~J.}\ \bibnamefont
  {Cha}},\ }\href@noop {} {\bibfield  {journal} {\bibinfo  {journal} {Nature Communications}\ }\textbf {\bibinfo {volume} {14}},\ \bibinfo {pages} {4803} (\bibinfo {year} {2023})}\BibitemShut {NoStop}%
\bibitem [{\citenamefont {Haim}\ \emph {et~al.}(2022)\citenamefont {Haim}, \citenamefont {Levchenko},\ and\ \citenamefont {Khodas}}]{haim2022mechanisms}%
  \BibitemOpen
  \bibfield  {author} {\bibinfo {author} {\bibfnamefont {M.}~\bibnamefont {Haim}}, \bibinfo {author} {\bibfnamefont {A.}~\bibnamefont {Levchenko}},\ and\ \bibinfo {author} {\bibfnamefont {M.}~\bibnamefont {Khodas}},\ }\href@noop {} {\bibfield  {journal} {\bibinfo  {journal} {Physical Review B}\ }\textbf {\bibinfo {volume} {105}},\ \bibinfo {pages} {024515} (\bibinfo {year} {2022})}\BibitemShut {NoStop}%
\bibitem [{\citenamefont {Yang}\ \emph {et~al.}(2018)\citenamefont {Yang}, \citenamefont {Wu},\ and\ \citenamefont {Li}}]{yang2018origin}%
  \BibitemOpen
  \bibfield  {author} {\bibinfo {author} {\bibfnamefont {Q.}~\bibnamefont {Yang}}, \bibinfo {author} {\bibfnamefont {M.}~\bibnamefont {Wu}},\ and\ \bibinfo {author} {\bibfnamefont {J.}~\bibnamefont {Li}},\ }\href@noop {} {\bibfield  {journal} {\bibinfo  {journal} {The Journal of Physical Chemistry Letters}\ }\textbf {\bibinfo {volume} {9}},\ \bibinfo {pages} {7160} (\bibinfo {year} {2018})}\BibitemShut {NoStop}%
\bibitem [{\citenamefont {Liu}\ \emph {et~al.}(2019)\citenamefont {Liu}, \citenamefont {Yang}, \citenamefont {Hu}, \citenamefont {Zhao}, \citenamefont {Chen},\ and\ \citenamefont {Ren}}]{liu2019vertical}%
  \BibitemOpen
  \bibfield  {author} {\bibinfo {author} {\bibfnamefont {X.}~\bibnamefont {Liu}}, \bibinfo {author} {\bibfnamefont {Y.}~\bibnamefont {Yang}}, \bibinfo {author} {\bibfnamefont {T.}~\bibnamefont {Hu}}, \bibinfo {author} {\bibfnamefont {G.}~\bibnamefont {Zhao}}, \bibinfo {author} {\bibfnamefont {C.}~\bibnamefont {Chen}},\ and\ \bibinfo {author} {\bibfnamefont {W.}~\bibnamefont {Ren}},\ }\href@noop {} {\bibfield  {journal} {\bibinfo  {journal} {Nanoscale}\ }\textbf {\bibinfo {volume} {11}},\ \bibinfo {pages} {18575} (\bibinfo {year} {2019})}\BibitemShut {NoStop}%
\end{thebibliography}%


\begin{thebibliography}{15}%
\makeatletter
\providecommand \@ifxundefined [1]{%
 \@ifx{#1\undefined}
}%
\providecommand \@ifnum [1]{%
 \ifnum #1\expandafter \@firstoftwo
 \else \expandafter \@secondoftwo
 \fi
}%
\providecommand \@ifx [1]{%
 \ifx #1\expandafter \@firstoftwo
 \else \expandafter \@secondoftwo
 \fi
}%
\providecommand \natexlab [1]{#1}%
\providecommand \enquote  [1]{``#1''}%
\providecommand \bibnamefont  [1]{#1}%
\providecommand \bibfnamefont [1]{#1}%
\providecommand \citenamefont [1]{#1}%
\providecommand \href@noop [0]{\@secondoftwo}%
\providecommand \href [0]{\begingroup \@sanitize@url \@href}%
\providecommand \@href[1]{\@@startlink{#1}\@@href}%
\providecommand \@@href[1]{\endgroup#1\@@endlink}%
\providecommand \@sanitize@url [0]{\catcode `\\12\catcode `\$12\catcode `\&12\catcode `\#12\catcode `\^12\catcode `\_12\catcode `\%12\relax}%
\providecommand \@@startlink[1]{}%
\providecommand \@@endlink[0]{}%
\providecommand \url  [0]{\begingroup\@sanitize@url \@url }%
\providecommand \@url [1]{\endgroup\@href {#1}{\urlprefix }}%
\providecommand \urlprefix  [0]{URL }%
\providecommand \Eprint [0]{\href }%
\providecommand \doibase [0]{https://doi.org/}%
\providecommand \selectlanguage [0]{\@gobble}%
\providecommand \bibinfo  [0]{\@secondoftwo}%
\providecommand \bibfield  [0]{\@secondoftwo}%
\providecommand \translation [1]{[#1]}%
\providecommand \BibitemOpen [0]{}%
\providecommand \bibitemStop [0]{}%
\providecommand \bibitemNoStop [0]{.\EOS\space}%
\providecommand \EOS [0]{\spacefactor3000\relax}%
\providecommand \BibitemShut  [1]{\csname bibitem#1\endcsname}%
\let\auto@bib@innerbib\@empty
%</preamble>
\bibitem [{\citenamefont {Wang}\ \emph {et~al.}(2013)\citenamefont {Wang}, \citenamefont {Meric}, \citenamefont {Huang}, \citenamefont {Gao}, \citenamefont {Gao}, \citenamefont {Tran}, \citenamefont {Taniguchi}, \citenamefont {Watanabe}, \citenamefont {Campos}, \citenamefont {Muller}, \citenamefont {Guo}, \citenamefont {Kim}, \citenamefont {Hone}, \citenamefont {Shepard},\ and\ \citenamefont {Dean}}]{wang2013one}%
  \BibitemOpen
  \bibfield  {author} {\bibinfo {author} {\bibfnamefont {L.}~\bibnamefont {Wang}}, \bibinfo {author} {\bibfnamefont {I.}~\bibnamefont {Meric}}, \bibinfo {author} {\bibfnamefont {P.}~\bibnamefont {Huang}}, \bibinfo {author} {\bibfnamefont {Q.}~\bibnamefont {Gao}}, \bibinfo {author} {\bibfnamefont {Y.}~\bibnamefont {Gao}}, \bibinfo {author} {\bibfnamefont {H.}~\bibnamefont {Tran}}, \bibinfo {author} {\bibfnamefont {T.}~\bibnamefont {Taniguchi}}, \bibinfo {author} {\bibfnamefont {K.}~\bibnamefont {Watanabe}}, \bibinfo {author} {\bibfnamefont {L.}~\bibnamefont {Campos}}, \bibinfo {author} {\bibfnamefont {D.}~\bibnamefont {Muller}}, \bibinfo {author} {\bibfnamefont {J.}~\bibnamefont {Guo}}, \bibinfo {author} {\bibfnamefont {P.}~\bibnamefont {Kim}}, \bibinfo {author} {\bibfnamefont {J.}~\bibnamefont {Hone}}, \bibinfo {author} {\bibfnamefont {K.}~\bibnamefont {Shepard}},\ and\ \bibinfo {author} {\bibfnamefont {C.}~\bibnamefont {Dean}},\ }\href@noop {} {\bibfield  {journal} {\bibinfo  {journal} {Science}\ }\textbf
  {\bibinfo {volume} {342}},\ \bibinfo {pages} {614} (\bibinfo {year} {2013})}\BibitemShut {NoStop}%
\bibitem [{\citenamefont {Schwartz}\ \emph {et~al.}(2019)\citenamefont {Schwartz}, \citenamefont {Chuang}, \citenamefont {Rosenberger}, \citenamefont {Sivaram}, \citenamefont {McCreary}, \citenamefont {Jonker},\ and\ \citenamefont {Centrone}}]{schwartz2019chemical}%
  \BibitemOpen
  \bibfield  {author} {\bibinfo {author} {\bibfnamefont {J.~J.}\ \bibnamefont {Schwartz}}, \bibinfo {author} {\bibfnamefont {H.-J.}\ \bibnamefont {Chuang}}, \bibinfo {author} {\bibfnamefont {M.~R.}\ \bibnamefont {Rosenberger}}, \bibinfo {author} {\bibfnamefont {S.~V.}\ \bibnamefont {Sivaram}}, \bibinfo {author} {\bibfnamefont {K.~M.}\ \bibnamefont {McCreary}}, \bibinfo {author} {\bibfnamefont {B.~T.}\ \bibnamefont {Jonker}},\ and\ \bibinfo {author} {\bibfnamefont {A.}~\bibnamefont {Centrone}},\ }\href@noop {} {\bibfield  {journal} {\bibinfo  {journal} {ACS Applied Materials \& Interfaces}\ }\textbf {\bibinfo {volume} {11}},\ \bibinfo {pages} {25578} (\bibinfo {year} {2019})}\BibitemShut {NoStop}%
\bibitem [{\citenamefont {Wan}\ \emph {et~al.}(2023)\citenamefont {Wan}, \citenamefont {Zheliuk}, \citenamefont {Yuan}, \citenamefont {Peng}, \citenamefont {Zhang}, \citenamefont {Liang}, \citenamefont {Zeitler}, \citenamefont {Wiedmann}, \citenamefont {Hussey}, \citenamefont {Palstra},\ and\ \citenamefont {Ye}}]{wan2023orbital}%
  \BibitemOpen
  \bibfield  {author} {\bibinfo {author} {\bibfnamefont {P.}~\bibnamefont {Wan}}, \bibinfo {author} {\bibfnamefont {O.}~\bibnamefont {Zheliuk}}, \bibinfo {author} {\bibfnamefont {N.~F.}\ \bibnamefont {Yuan}}, \bibinfo {author} {\bibfnamefont {X.}~\bibnamefont {Peng}}, \bibinfo {author} {\bibfnamefont {L.}~\bibnamefont {Zhang}}, \bibinfo {author} {\bibfnamefont {M.}~\bibnamefont {Liang}}, \bibinfo {author} {\bibfnamefont {U.}~\bibnamefont {Zeitler}}, \bibinfo {author} {\bibfnamefont {S.}~\bibnamefont {Wiedmann}}, \bibinfo {author} {\bibfnamefont {N.~E.}\ \bibnamefont {Hussey}}, \bibinfo {author} {\bibfnamefont {T.~T.}\ \bibnamefont {Palstra}},\ and\ \bibinfo {author} {\bibfnamefont {J.}~\bibnamefont {Ye}},\ }\href@noop {} {\bibfield  {journal} {\bibinfo  {journal} {Nature}\ }\textbf {\bibinfo {volume} {619}},\ \bibinfo {pages} {46} (\bibinfo {year} {2023})}\BibitemShut {NoStop}%
\bibitem [{\citenamefont {Fatemi}\ \emph {et~al.}(2018)\citenamefont {Fatemi}, \citenamefont {Wu}, \citenamefont {Cao}, \citenamefont {Bretheau}, \citenamefont {Gibson}, \citenamefont {Watanabe}, \citenamefont {Taniguchi}, \citenamefont {Cava},\ and\ \citenamefont {Jarillo-Herrero}}]{fatemi2018electrically}%
  \BibitemOpen
  \bibfield  {author} {\bibinfo {author} {\bibfnamefont {V.}~\bibnamefont {Fatemi}}, \bibinfo {author} {\bibfnamefont {S.}~\bibnamefont {Wu}}, \bibinfo {author} {\bibfnamefont {Y.}~\bibnamefont {Cao}}, \bibinfo {author} {\bibfnamefont {L.}~\bibnamefont {Bretheau}}, \bibinfo {author} {\bibfnamefont {Q.~D.}\ \bibnamefont {Gibson}}, \bibinfo {author} {\bibfnamefont {K.}~\bibnamefont {Watanabe}}, \bibinfo {author} {\bibfnamefont {T.}~\bibnamefont {Taniguchi}}, \bibinfo {author} {\bibfnamefont {R.~J.}\ \bibnamefont {Cava}},\ and\ \bibinfo {author} {\bibfnamefont {P.}~\bibnamefont {Jarillo-Herrero}},\ }\href@noop {} {\bibfield  {journal} {\bibinfo  {journal} {Science}\ }\textbf {\bibinfo {volume} {362}},\ \bibinfo {pages} {926} (\bibinfo {year} {2018})}\BibitemShut {NoStop}%
\bibitem [{\citenamefont {Yankowitz}\ \emph {et~al.}(2019)\citenamefont {Yankowitz}, \citenamefont {Chen}, \citenamefont {Polshyn}, \citenamefont {Zhang}, \citenamefont {Watanabe}, \citenamefont {Taniguchi}, \citenamefont {Graf}, \citenamefont {Young},\ and\ \citenamefont {Dean}}]{yankowitz2019tuning}%
  \BibitemOpen
  \bibfield  {author} {\bibinfo {author} {\bibfnamefont {M.}~\bibnamefont {Yankowitz}}, \bibinfo {author} {\bibfnamefont {S.}~\bibnamefont {Chen}}, \bibinfo {author} {\bibfnamefont {H.}~\bibnamefont {Polshyn}}, \bibinfo {author} {\bibfnamefont {Y.}~\bibnamefont {Zhang}}, \bibinfo {author} {\bibfnamefont {K.}~\bibnamefont {Watanabe}}, \bibinfo {author} {\bibfnamefont {T.}~\bibnamefont {Taniguchi}}, \bibinfo {author} {\bibfnamefont {D.}~\bibnamefont {Graf}}, \bibinfo {author} {\bibfnamefont {A.~F.}\ \bibnamefont {Young}},\ and\ \bibinfo {author} {\bibfnamefont {C.~R.}\ \bibnamefont {Dean}},\ }\href@noop {} {\bibfield  {journal} {\bibinfo  {journal} {Science}\ }\textbf {\bibinfo {volume} {363}},\ \bibinfo {pages} {1059} (\bibinfo {year} {2019})}\BibitemShut {NoStop}%
\bibitem [{\citenamefont {Hohenberg}\ and\ \citenamefont {Kohn}(1964)}]{hohenkohn1964}%
  \BibitemOpen
  \bibfield  {author} {\bibinfo {author} {\bibfnamefont {P.}~\bibnamefont {Hohenberg}}\ and\ \bibinfo {author} {\bibfnamefont {W.}~\bibnamefont {Kohn}},\ }\href {https://doi.org/10.1103/PhysRev.136.B864} {\bibfield  {journal} {\bibinfo  {journal} {Phys. Rev.}\ }\textbf {\bibinfo {volume} {136}},\ \bibinfo {pages} {B864} (\bibinfo {year} {1964})}\BibitemShut {NoStop}%
\bibitem [{\citenamefont {Kohn}\ and\ \citenamefont {Sham}(1965)}]{kohnsham1965}%
  \BibitemOpen
  \bibfield  {author} {\bibinfo {author} {\bibfnamefont {W.}~\bibnamefont {Kohn}}\ and\ \bibinfo {author} {\bibfnamefont {L.~J.}\ \bibnamefont {Sham}},\ }\href {https://doi.org/10.1103/PhysRev.140.A1133} {\bibfield  {journal} {\bibinfo  {journal} {Phys. Rev.}\ }\textbf {\bibinfo {volume} {140}},\ \bibinfo {pages} {A1133} (\bibinfo {year} {1965})}\BibitemShut {NoStop}%
\bibitem [{\citenamefont {Kresse}\ and\ \citenamefont {Furthm\"uller}(1996)}]{vasp1996prb}%
  \BibitemOpen
  \bibfield  {author} {\bibinfo {author} {\bibfnamefont {G.}~\bibnamefont {Kresse}}\ and\ \bibinfo {author} {\bibfnamefont {J.}~\bibnamefont {Furthm\"uller}},\ }\href {https://doi.org/10.1103/PhysRevB.54.11169} {\bibfield  {journal} {\bibinfo  {journal} {Phys. Rev. B}\ }\textbf {\bibinfo {volume} {54}},\ \bibinfo {pages} {11169} (\bibinfo {year} {1996})}\BibitemShut {NoStop}%
\bibitem [{\citenamefont {Kresse}\ and\ \citenamefont {Furthmüller}(1996)}]{vasp1996cms}%
  \BibitemOpen
  \bibfield  {author} {\bibinfo {author} {\bibfnamefont {G.}~\bibnamefont {Kresse}}\ and\ \bibinfo {author} {\bibfnamefont {J.}~\bibnamefont {Furthmüller}},\ }\href {https://doi.org/https://doi.org/10.1016/0927-0256(96)00008-0} {\bibfield  {journal} {\bibinfo  {journal} {Computational Materials Science}\ }\textbf {\bibinfo {volume} {6}},\ \bibinfo {pages} {15} (\bibinfo {year} {1996})}\BibitemShut {NoStop}%
\bibitem [{\citenamefont {Blöchl}(1994)}]{blochl1994paw}%
  \BibitemOpen
  \bibfield  {author} {\bibinfo {author} {\bibfnamefont {P.~E.}\ \bibnamefont {Blöchl}},\ }\href {https://doi.org/10.1103/PhysRevB.50.17953} {\bibfield  {journal} {\bibinfo  {journal} {Physical Review B}\ }\textbf {\bibinfo {volume} {50}},\ \bibinfo {pages} {17953} (\bibinfo {year} {1994})}\BibitemShut {NoStop}%
\bibitem [{\citenamefont {Perdew}\ \emph {et~al.}(1996)\citenamefont {Perdew}, \citenamefont {Burke},\ and\ \citenamefont {Ernzerhof}}]{pbe1996gga}%
  \BibitemOpen
  \bibfield  {author} {\bibinfo {author} {\bibfnamefont {J.~P.}\ \bibnamefont {Perdew}}, \bibinfo {author} {\bibfnamefont {K.}~\bibnamefont {Burke}},\ and\ \bibinfo {author} {\bibfnamefont {M.}~\bibnamefont {Ernzerhof}},\ }\href {https://doi.org/10.1103/PhysRevLett.77.3865} {\bibfield  {journal} {\bibinfo  {journal} {Physical Review Letters}\ }\textbf {\bibinfo {volume} {77}},\ \bibinfo {pages} {3865} (\bibinfo {year} {1996})}\BibitemShut {NoStop}%
\bibitem [{\citenamefont {Ong}\ \emph {et~al.}(2013)\citenamefont {Ong}, \citenamefont {Richards}, \citenamefont {Jain}, \citenamefont {Hautier}, \citenamefont {Kocher}, \citenamefont {Cholia}, \citenamefont {Gunter}, \citenamefont {Chevrier}, \citenamefont {Persson},\ and\ \citenamefont {Ceder}}]{ong2013pymatgen}%
  \BibitemOpen
  \bibfield  {author} {\bibinfo {author} {\bibfnamefont {S.~P.}\ \bibnamefont {Ong}}, \bibinfo {author} {\bibfnamefont {W.~D.}\ \bibnamefont {Richards}}, \bibinfo {author} {\bibfnamefont {A.}~\bibnamefont {Jain}}, \bibinfo {author} {\bibfnamefont {G.}~\bibnamefont {Hautier}}, \bibinfo {author} {\bibfnamefont {M.}~\bibnamefont {Kocher}}, \bibinfo {author} {\bibfnamefont {S.}~\bibnamefont {Cholia}}, \bibinfo {author} {\bibfnamefont {D.}~\bibnamefont {Gunter}}, \bibinfo {author} {\bibfnamefont {V.~L.}\ \bibnamefont {Chevrier}}, \bibinfo {author} {\bibfnamefont {K.~A.}\ \bibnamefont {Persson}},\ and\ \bibinfo {author} {\bibfnamefont {G.}~\bibnamefont {Ceder}},\ }\href@noop {} {\bibfield  {journal} {\bibinfo  {journal} {Computational Materials Science}\ }\textbf {\bibinfo {volume} {68}},\ \bibinfo {pages} {314} (\bibinfo {year} {2013})}\BibitemShut {NoStop}%
\bibitem [{\citenamefont {Herath}\ \emph {et~al.}(2020)\citenamefont {Herath}, \citenamefont {Tavadze}, \citenamefont {He}, \citenamefont {Bousquet}, \citenamefont {Singh}, \citenamefont {Muñoz},\ and\ \citenamefont {Romero}}]{PyProcar2020}%
  \BibitemOpen
  \bibfield  {author} {\bibinfo {author} {\bibfnamefont {U.}~\bibnamefont {Herath}}, \bibinfo {author} {\bibfnamefont {P.}~\bibnamefont {Tavadze}}, \bibinfo {author} {\bibfnamefont {X.}~\bibnamefont {He}}, \bibinfo {author} {\bibfnamefont {E.}~\bibnamefont {Bousquet}}, \bibinfo {author} {\bibfnamefont {S.}~\bibnamefont {Singh}}, \bibinfo {author} {\bibfnamefont {F.}~\bibnamefont {Muñoz}},\ and\ \bibinfo {author} {\bibfnamefont {A.~H.}\ \bibnamefont {Romero}},\ }\href {https://doi.org/https://doi.org/10.1016/j.cpc.2019.107080} {\bibfield  {journal} {\bibinfo  {journal} {Computer Physics Communications}\ }\textbf {\bibinfo {volume} {251}},\ \bibinfo {pages} {107080} (\bibinfo {year} {2020})}\BibitemShut {NoStop}%
\bibitem [{\citenamefont {Lang}\ \emph {et~al.}(2024)\citenamefont {Lang}, \citenamefont {Tavadze}, \citenamefont {Tellez}, \citenamefont {Bousquet}, \citenamefont {Xu}, \citenamefont {Muñoz}, \citenamefont {Vasquez}, \citenamefont {Herath},\ and\ \citenamefont {Romero}}]{PyProcar2024}%
  \BibitemOpen
  \bibfield  {author} {\bibinfo {author} {\bibfnamefont {L.}~\bibnamefont {Lang}}, \bibinfo {author} {\bibfnamefont {P.}~\bibnamefont {Tavadze}}, \bibinfo {author} {\bibfnamefont {A.}~\bibnamefont {Tellez}}, \bibinfo {author} {\bibfnamefont {E.}~\bibnamefont {Bousquet}}, \bibinfo {author} {\bibfnamefont {H.}~\bibnamefont {Xu}}, \bibinfo {author} {\bibfnamefont {F.}~\bibnamefont {Muñoz}}, \bibinfo {author} {\bibfnamefont {N.}~\bibnamefont {Vasquez}}, \bibinfo {author} {\bibfnamefont {U.}~\bibnamefont {Herath}},\ and\ \bibinfo {author} {\bibfnamefont {A.~H.}\ \bibnamefont {Romero}},\ }\href {https://doi.org/https://doi.org/10.1016/j.cpc.2023.109063} {\bibfield  {journal} {\bibinfo  {journal} {Computer Physics Communications}\ }\textbf {\bibinfo {volume} {297}},\ \bibinfo {pages} {109063} (\bibinfo {year} {2024})}\BibitemShut {NoStop}%
\bibitem [{\citenamefont {Fukui}\ and\ \citenamefont {Hatsugai}(2007)}]{Fukui2007nfield}%
  \BibitemOpen
  \bibfield  {author} {\bibinfo {author} {\bibfnamefont {T.}~\bibnamefont {Fukui}}\ and\ \bibinfo {author} {\bibfnamefont {Y.}~\bibnamefont {Hatsugai}},\ }\href {https://doi.org/10.1143/JPSJ.76.053702} {\bibfield  {journal} {\bibinfo  {journal} {Journal of the Physical Society of Japan}\ }\textbf {\bibinfo {volume} {76}},\ \bibinfo {pages} {053702} (\bibinfo {year} {2007})}\BibitemShut {NoStop}%
\end{thebibliography}%
\end{document}

%%%%%%%%%%%%%%%%%%%%%%%%%%%%%%%%%%%%%%%%%%%%%%%%%%%%%%%%%%%%%%%%%%%%%%%%%%%
%%%%%%%%%%%%%%%%%%%%%%%%%%%%%%%%%%%%%%%%%%%%%%%%%%%%%%%%%%%%%%%%%%%%%%%%%%%
%%%%%%%%%%%%%%%%%%%%%%%%%%%%%%%%%%%%%%%%%%%%%%%%%%%%%%%%%%%%%%%%%%%%%%%%%%%
% AUTHORS
\author{Zizhong Li}
\thanks{These two authors contributed equally}
\affiliation{Department of Materials Science and Engineering, University of Wisconsin-Madison, Madison, WI, 53706}
\author{Apoorv Jindal}
\thanks{These two authors contributed equally}
\affiliation{Department of Physics, Columbia University, New York, NY, 10027}
\author{Alex Strasser}
\affiliation{Department of Materials Science and Engineering, Texas A$\&$M University, College Station, TX, 77843}
\author{Yangchen He}
\affiliation{Department of Materials Science and Engineering, University of Wisconsin-Madison, Madison, WI, 53706}
\author{Wenkai Zheng}
\affiliation{Department of Physics, Florida State University, Tallahassee, FL, 32306}
\affiliation{National High Magnetic Field Laboratory, Tallahassee, FL, 32310}
\author{David Graf}
\affiliation{National High Magnetic Field Laboratory, Tallahassee, FL, 32310}
\author{Takashi Taniguchi}
\affiliation{National Institute for Materials Science, Tsukuba, Japan}
\author{Kenji Watanabe}
\affiliation{National Institute for Materials Science, Tsukuba, Japan}
\author{Luis Balicas}
\affiliation{National High Magnetic Field Laboratory, Tallahassee, FL, 32310}
\author{Cory R. Dean}
\affiliation{Department of Physics, Columbia University, New York, NY, 10027}
\author{Xiaofeng Qian}
\email{feng@tamu.edu}
\affiliation{Department of Materials Science and Engineering, Texas A$\&$M University, College Station, TX, 77843}
\author{Abhay N. Pasupathy}
\email{apn2108@columbia.edu}
\affiliation{Department of Physics, Columbia University, New York, NY, 10027}
\author{Daniel A. Rhodes}
\email{darhodes@wisc.edu}
\affiliation{Department of Materials Science and Engineering, University of Wisconsin-Madison, Madison, WI, 53706}

\title{\vspace*{0.3in} Supplemental Material \\ \vspace{0.3in}  Two-Fold Anisotropic Superconductivity in Bilayer $T_d$-MoTe$_2$}
\maketitle
\begin{figure}[t] 
    %\centering
	\includegraphics[width=\linewidth]{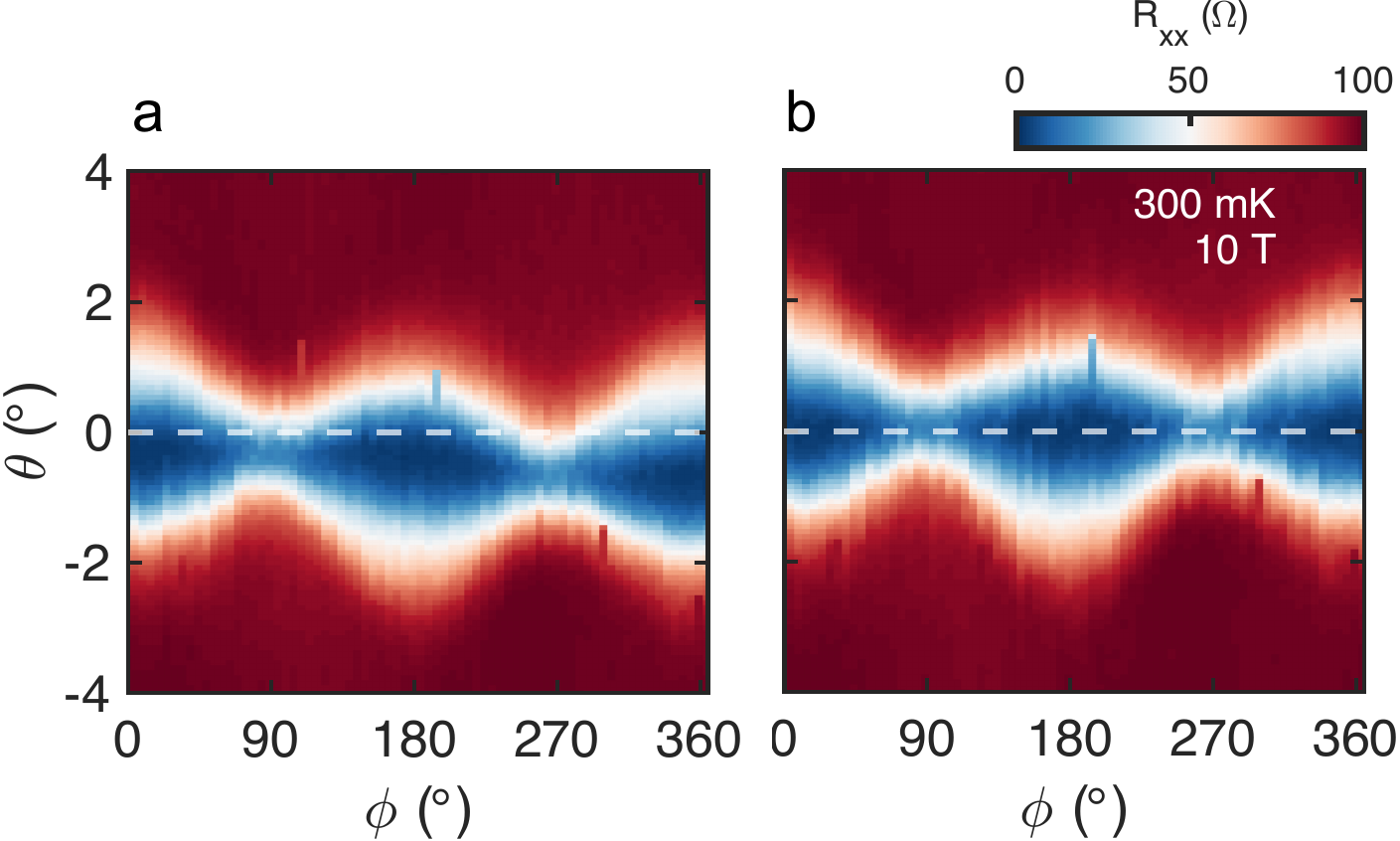}
    \caption{Eliminating canting error in angular mapping. (a,b) Angular mapping of the longitude resistance in bilayer MoTe$_2$, with (a) the original data and (b) mapping after subtracting the effect of canting angle.} 
\label{S1}
\end{figure}
\begin{figure}[t] 
    \centering
    \includegraphics[width=\linewidth]{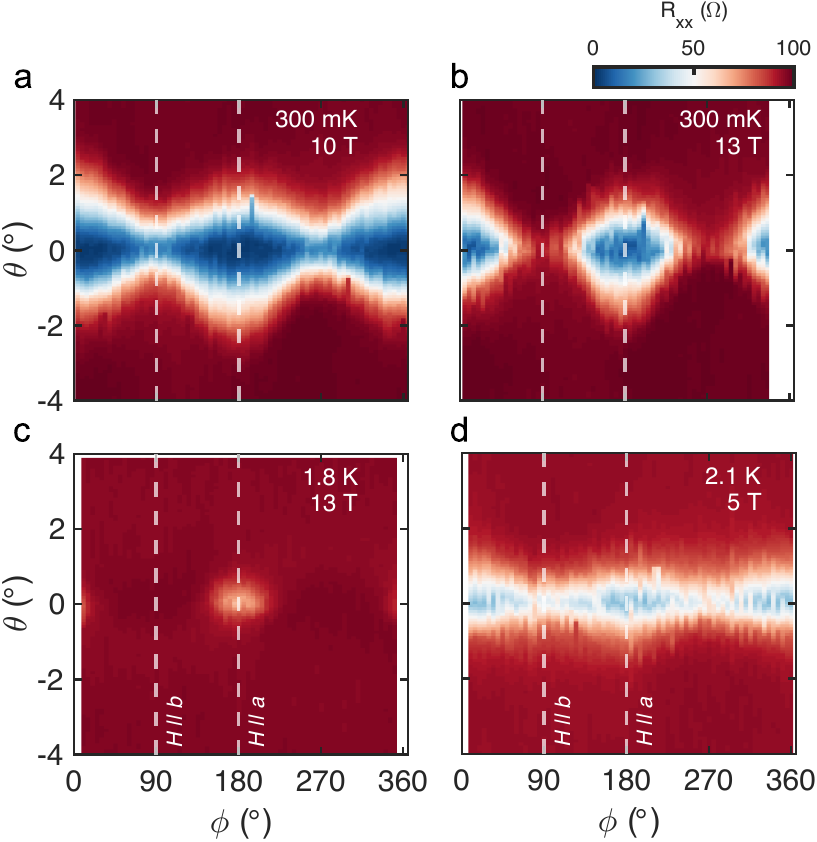}
    \caption{Angular mapping of resistance $R_{xx}$ at different temperature and magnetic field. (a) 300 mK, 10 T. (b) 300 mK, 13 T. (c) 1.8 K, 13 T. (d) 2.1 K, 5 T. The two-fold symmetry preserves at all temperatures and magnetic fields.}	
\label{S2}
\end{figure}
\begin{figure}[t] 
    \centering
    \includegraphics[width=\linewidth]{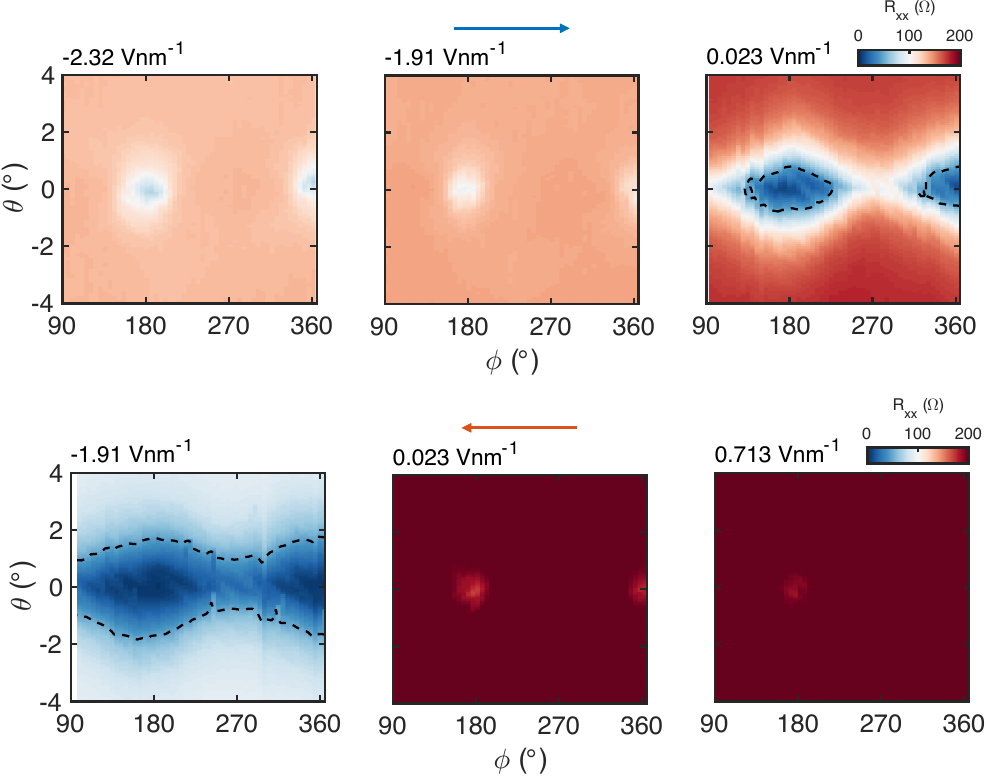}
    \caption{Angular mapping of $R_{xx}$ at different stages in ferroelectric switching. (a-c) Maps with up polarization. (d-f) After ferroelectric switching, maps with down polarization. }
\label{S3}
\end{figure}
\begin{figure}[t] 
    \centering
    \includegraphics[width=\linewidth]{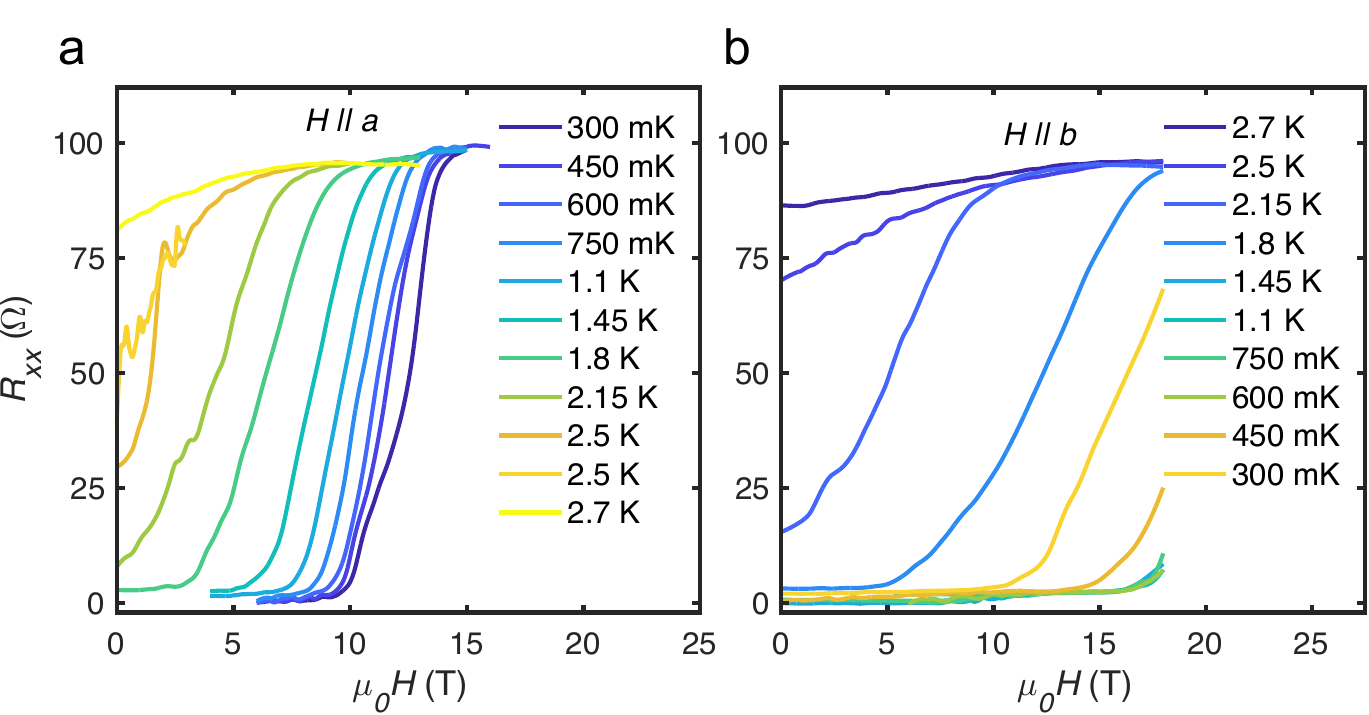}
    \caption{The suppression of superconductivity with magnetic fields. (a) $R_{xx}$ versus magnetic field when field parallel to crystal $a$-axis, the number in legends are the ratios $T/T_c$. (b) $R_{xx}$ versus magnetic field when field parallel to crystal $b$-axis.}
\label{S4}
\end{figure}
\begin{figure}[t] 
    \centering
    \includegraphics[width=\linewidth]{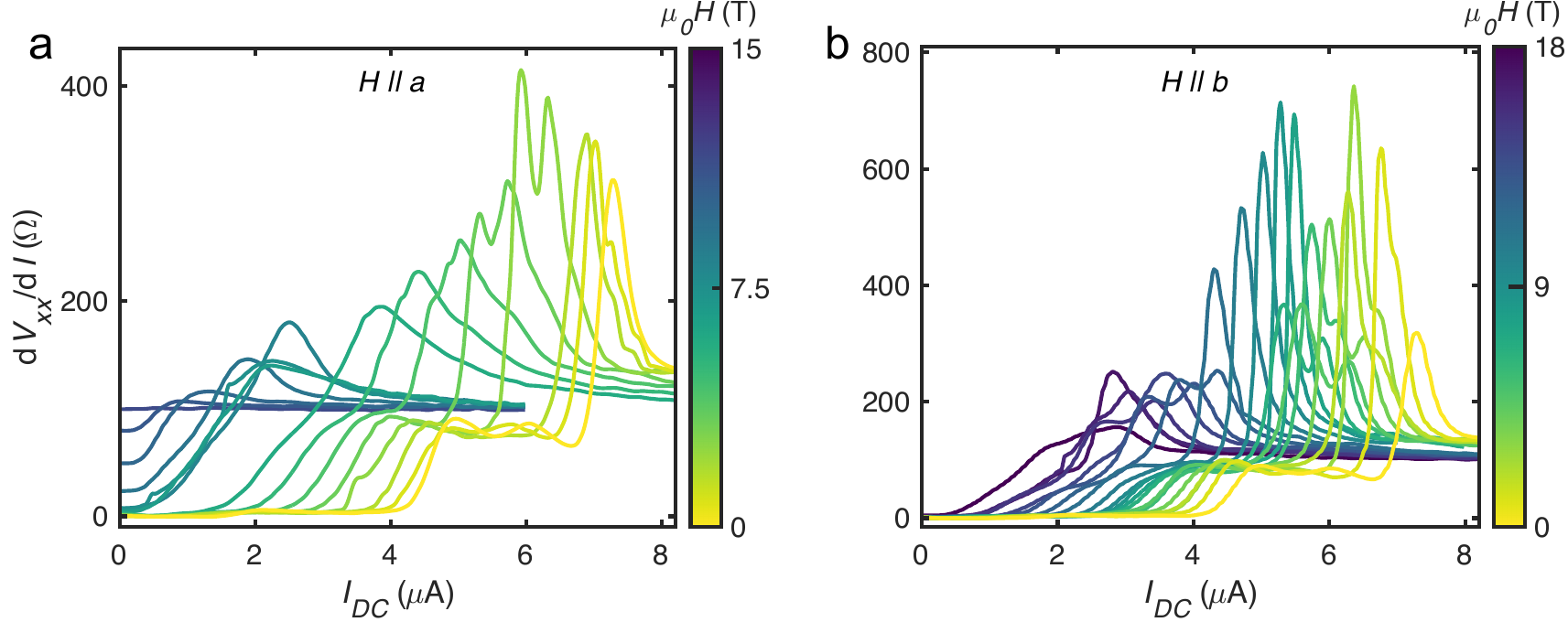}
    \caption{Critical current measurements. (a) Differential resistance versus DC bias when field parallel to $a$-axis. Multiple coherence peaks arise may indicate the existence of multiple asynchronous superconducting transitions. (b) Differential resistance versus DC bias when field parallel to $b$-axis.}
\label{S5}
\end{figure}

\newpage
%% Introduction
\noindent \textbf{\large1. Single Crystal Growth}\vspace{2.5mm}\\
%%%%%%%%%%%%%%%%%%%%%%%%%%%%%%%%%%%%%%%%%%%%%%%%%%%%%%%%%%%%%%%%%%
\noindent Single crystals were prepared by combining Mo powder (99.997\%, Fisher Scientific AA3968614) and Te lumps (99.9999\%, Fisher Scientific AA1075830) in a ratio of 1:20 inside of a Canfield crucible (LSP Ceramics), which was subsequently sealed under vacuum ($\sim 5 \times 10^{-6}$ Torr) inside of a quartz ampoule. The reagents were then heated to 1120 $^{\circ}$C within 24 h and held at this temperature for 1 week before cooling to 880 $^{\circ}$C at a rate of 0.5 $^{\circ}$C/h. At 880 $^{\circ}$C the excess Te flux was decanted in a centrifuge, quenching the samples in air. The subsequently harvested 1T$^\prime$-MoTe$_2$ single crystals were sealed in a second quartz ampoule under vacuum and annealed at 425 $^{\circ}$C with a 200 $^{\circ}$C gradient for 48 h to remove any residual tellurium. A typical single crystal synthesized using this process has residual resistivity ratios ranging from 600 to 2000.\vspace{2.5mm}\\  
%%%%%%%%%%%%%%%%%%%%%%%%%%%%%%%%%%%%%%%%%%%%%%%%%%%%%%%%%%%%%%%%%%
\textbf{\large2. Heterostructure Fabrication}\vspace{2.5mm}\\
%%%%%%%%%%%%%%%%%%%%%%%%%%%%%%%%%%%%%%%%%%%%%%%%%%%%%%%%%%%%%%%%%%
\noindent The dual gate device is fabricated in three stages of processing. For the first stage, metal backgates were deposited onto 285 nm SiO$_2$ using standard e-beam lithography (EBL) techniques and a bilayer resist of PMMA 495A4/PMMA 950A2. After patterning, 10 nm of Pd was deposited via e-beam deposition, followed by annealing in vacuum at 300 $^\circ$C, and a low power O$_2$ plasma for 10 minutes to remove any residual scum on the prepatterned Pd backgates. Afterwards, 14-30 nm of hexagonal boron nitride ($h$-BN) was transferred onto the Pd backgates using a dry stacking method~\cite{wang2013one} with either poly(bisphenol a carbonate), PC, or poly(propylene carbonate), PPC, as the stacking polymer. After transfer, the $h$-BN+Pd backgate was rinsed in either chloroform (PC) or acetone (PPC), followed by IPA and deionized water. Contacts were then patterned onto the $h$-BN using standard EBL, followed by e-beam deposition of Au (12 nm). Liftoff was performed in acetone for 30 minutes. Finally, prepatterned contacts were cleaned by atomic force microscopy (AFM) using an Asylum Research Cypher-S in contact mode with an AC-160 AFM tip with a force of $\sim 500$ nN. In the second stage, another $h$-BN was picked up using PPC to act as the $h$-BN for the top gate. Bilayer MoTe$_2$ was then exfoliated inside of a glovebox ($<$1 ppm H$_2$O/O$_2$) by using toluene-cleaned PDMS (soaked for three days)~\cite{schwartz2019chemical} and transferring onto 285 nm SiO$_2$. Next, the bilayer was picked up by the PPC/$h$-BN stamp and then immediately stacked onto the prepatterned contacts. Careful precaution is taken to make sure the prepatterned contacts are completely encapsulated in the top and bottom $h$-BN layers to any degradation while the sample is exposed to air during subsequent steps and measurements. In the final stage, etching windows were patterned on the encapsulated contacts using EBL, and etched using a CHF$_3$/O$_2$ plasma mixed. The resist was then stripped and the bonding pads patterned onto the exposed prepatterned contacts along with the top gate contact via EBL. Finally, 5 nm/15 nm/50 nm Ti/Pd/Au was deposited using e-beam deposition. The device was additionally covered with a PMMA 950A2 mask to further protect against oxidation. We have found that, when using these procedures in combination with high-quality single crystals, device quality is primarily limited by the purity of the glovebox.\vspace{2.5mm}\\
%%%%%%%%%%%%%%%%%%%%%%%%%%%%%%%%%%%%%%%%%%%%%%%%%%%%%%%%%%%%%%%%%%
\textbf{\large3. Experimental Setup}\vspace{2.5mm}\\
%%%%%%%%%%%%%%%%%%%%%%%%%%%%%%%%%%%%%%%%%%%%%%%%%%%%%%%%%%%%%%%%%%
\noindent\textbf{Aligning Samples in Magnetic Field:}\\
\noindent The in-plane anisotropy measurement in MoTe$_2$ is extremely susceptible to magnetic field canting, due to large anisotropy in the out-of-plane magnetoresistance~\cite{wan2023orbital}. Therefore, careful alignment of the sample to a parallel position is needed for accurate measurement. We therefore used a combination of a piezo-rotator, in-plane $\phi$ angle, and spring rotator, $\theta$ angle, designed by the National High Magnetic Field Laboratory to three-dimensionally align the sample and ensure the field is parallel to the plane of the sample at all in-plane angles explored. Before measurement, we first aligned the devices coarsely by finding the zero points of the Hall sensors that are aligned parallel to the crystal $a-c$ and $b-c$ planes, and maximizing the value of the Hall sensor parallel to the a-b plane. Subsequently, we finely aligned the devices by finding minimizing the longitudinal resistance of the bilayer MoTe$_2$ device in magnetic field.\\
During the angular measurements, we ran two loops, an inner loop where the spring rotator was swept back and forth from -7 $^{\circ}$ to 7 $^{\circ}$, and a slower outer loop where the piezo-rotator was ramped from 0 $^{\circ}$ to 360 $^{\circ}$. Due to the backlash of the spring rotator, only measurement results in the forward direction between -4 $^{\circ}$ and 4 $^{\circ}$ were taken to construct maps. An example angular measurement result is shown in Fig. \ref{S1}a. Here, we can see the canting angle along the $\theta$-direction (or out-of-plane direction) is within 1 $^{\circ}$ throughout the whole map. For bilayer MoTe$_2$, we expect it to be isotropic along the out-of-plane direction and so any offsets from perfect symmetry about $\theta$ is attributed to a canting angle, which we subtract out by polynomial fitting along the minimum resistance points in the map, as shown in Fig. \ref{S1}b). Additional angular maps are given for different temperature and fields (Fig. \ref{S2}) and along the ferroelectric hysteresis loop (Fig. \ref{S3})\\
\indent For all figures where we plot the angular mapping in resistance with different displacement fields, $D$, and doping concentrations, $\Delta n$, the calculation of doping carrier density ${\Delta}n$ and displacement $D$ in our dual-gate device follows the geometry equations: ${\Delta}n={\epsilon}_{h-BN}{\epsilon}_0(V_\textrm{T}/d_T+V_B/d_B)/e$, and $D={\epsilon}_{h-BN}(V_T/d_T-V_B/d_B)/2$. Here, ${\epsilon}_{h-BN}$ is the dielectric constant of $h$-BN, ${\epsilon}_0$ is the vacuum permittivity, $d_T$ and $d_B$ are the thickness of top and bottom $h$-BN, and $e$ is the charge of an electron.\vspace{2.5mm}\\
%%%%%%%%%%%%%%%%%%%%%%%%%%%%%%%%%%%%%%%%%%%%%%%%%%%%%%%%%%%%%%%%%%
\noindent\textbf{Transport Measurements:}\\
%%%%%%%%%%%%%%%%%%%%%%%%%%%%%%%%%%%%%%%%%%%%%%%%%%%%%%%%%%%%%%%%%%
\noindent From the main text Fig. 2, we plotted the $H_{c2}-T$ and $I_c-B$ relationship. The values used for critical field ($H_\textrm{c2}$) versus $T$ and for the critical current ($I_\textrm{c})$ versus $H_\textrm{c2}$ presented in the main text were extracted from the measurements presented in Fig. \ref{S4} and Fig. \ref{S5}, respectively. To extract the $T$-dependence of $H_\textrm{c2}$, we measured the field-dependent superconducting transition at fixed $T$, extracting the values of $H_\textrm{c2}$ for 90\%, 50\%, and 10\% of the normal state resistance ($\sim 100$ $\Omega$). Notably, in Fig. \ref{S4} (left panel), the superconducting transition curve manifests slight fluctuations in $R_{xx}$ at temperatures near 2 K, which are induced by the temperature controller due to the difficulty in maintaining good helium flow for establishing a stable PID. Therefore, data points near 2 K were excluded from the $H_{c2}-T$ fittings given in Fig. 2 of the main text. To acquire a $I_\textrm{c}-B$ diagram, we measured the differential resistance ($dV_\textrm{xx}/dI$) versus DC bias at fixed $B$, where the maximum resistance peak was denoted as the critical current coherence peak. Like many other 2D superconductors~\cite{fatemi2018electrically, yankowitz2019tuning}, we observe sub-peaks/shoulders arising before the highest coherence peak. This indicates that domains or disorder may exist in our device. The most likely cause of this is inhomogeneous strain due to the pre-patterned contact strategy that we employ to successfully avoid air degradation.\vspace{2.5mm}\\
%%%%%%%%%%%%%%%%%%%%%%%%%%%%%%%%%%%%%%%%%%%%%%%%%%%%%%%%%%%%%%%%%%%%%%%%%%%%%%%%%%%%%%%%%%%%%%%%%%%%%%%%%%%%%%%%%%%%%%
\textbf{\large4. Theoretical Calculations}\vspace{2.5mm}\\
%%%%%%%%%%%%%%%%%%%%%%%%%%%%%%%%%%%%%%%%%%%%%%%%%%%%%%%%%
First-principles density functional theory calculations~\cite{hohenkohn1964,kohnsham1965} were performed using the Vienna Ab initio Simulation Package (VASP)~\cite{vasp1996prb,vasp1996cms} with the projector augmented wave method (PAW) method~\cite{blochl1994paw}, the Perdew-Burke-Ernzerhof (PBE) exchange-correlation energy functional~\cite{pbe1996gga}, and a plane-wave basis with an energy cutoff of 400 eV. The initial bilayer $T_d$-MoTe$_2$ structure was obtained from the bulk crystal structure. Subsequently, a vacuum layer of $\sim$20 \AA~was added along the out-of-plane (\textit{z})-direction (i.e. along the $c$-axis) of the bilayer structure to minimize the periodic image interactions. The crystal structures, including both atomic positions and in-plane lattice vectors, were then relaxed with a maximum residual force of 0.01 eV/\AA, a total energy convergence criterion of $10^{-6}$ eV, a smearing factor of 0.13 eV, and a Monkhorst-Pack \textit{\textbf{k}}-point sampling of $20\times10\times1$. Both spin-orbit coupling and dipole corrections were included in the self-consistent electronic and band structure calculations. The resulting band structures were plotted using \textit{pymatgen}~\cite{ong2013pymatgen}. The spin textures were calculated and plotted using \textit{PyProcar}~\cite{PyProcar2020,PyProcar2024} using a \textit{\textbf{k}}-point grid of $30\times30$ from -0.25 to 0.25 in fractional coordinates along $k_x$ and $k_y$ at $E=E_F-0.01$ eV in order to better visualize the spin texture around the electron and hole pockets. Furthermore, electric field was applied along the out-of-plane direction to study the field-dependent electronic band structure and spin texture. Finally, since ferroelectric bilayer MoTe$_2$ is noncentrosymmetric, the Z$_2$ topological invariant was calculated by using the $n$-field method~\cite{Fukui2007nfield}, and the result confirms that bilayer MoTe$_2$ has a Z$_2$ topological invariant of 0 and is thus a Z$_2$-trivial system.
\newpage
%%%%%%%%%%%%%%%%%%% S-FIGURE 1 %%%%%%%%%%%%%%%
\clearpage
%%%%%%%%%%%%%%%%%%% S-FIGURE 2 %%%%%%%%%%%%%%
\clearpage
%%%%%%%%%%%%%%%%%%% S-FIGURE 3 %%%%%%%%%%%%%%
%
\clearpage
%%%%%%%%%%%%%%%%%%% S-FIGURE 4 %%%%%%%%%%%%%%
%
\clearpage
%%%%%%%%%%%%%%%%%%% S-FIGURE 5 %%%%%%%%%%%%%%
%
\clearpage

%%%%%%%%%%%%%%%%%%% S-FIGURE 6 %%%%%%%%%%%%%%
%
\begin{figure*}[t] 
    \centering
	\includegraphics[width=\linewidth]{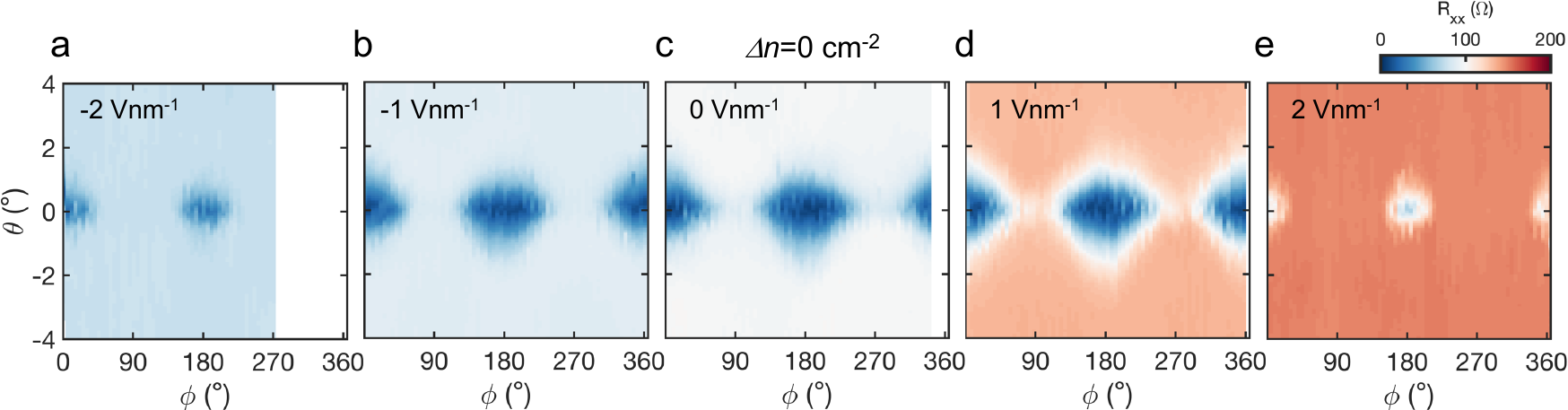}
	\caption{Angular mapping of resistance at different displacement field. The two-fold symmetry is robust against high displacement field.}
	
\label{S6}
\end{figure*}
\clearpage

%%%%%%%%%%%%%%%%%%% S-FIGURE 7 %%%%%%%%%%%%%
%
\begin{figure*}[t] 
    \centering
	\includegraphics[width=\linewidth]{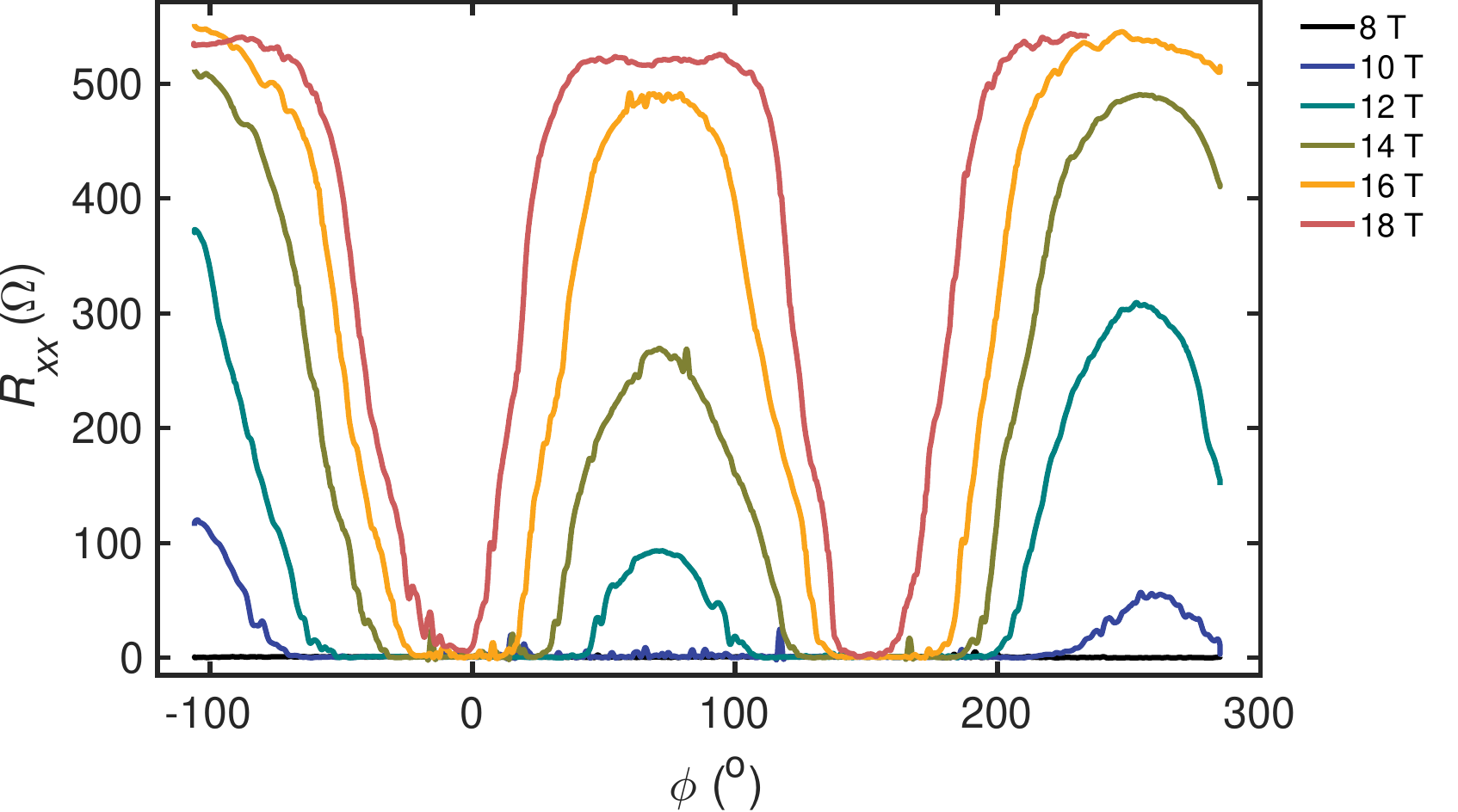}
	\caption{In-plane anisotropy in sample S2. Resistance vs in-plane angle $\phi$ and for several values of magnetic field applied to sample S2 at $T$ = 300 mK. The deviation of superconducting dip from the crystal principal axis results from magnetic field canting.}
	
\label{S7}
\end{figure*}
\clearpage

%%%%%%%%%%%%%%%%%%% S-FIGURE 8 %%%%%%%%%%%%%
%
\begin{figure*}[t] 
    \centering
	\includegraphics[width=\linewidth]{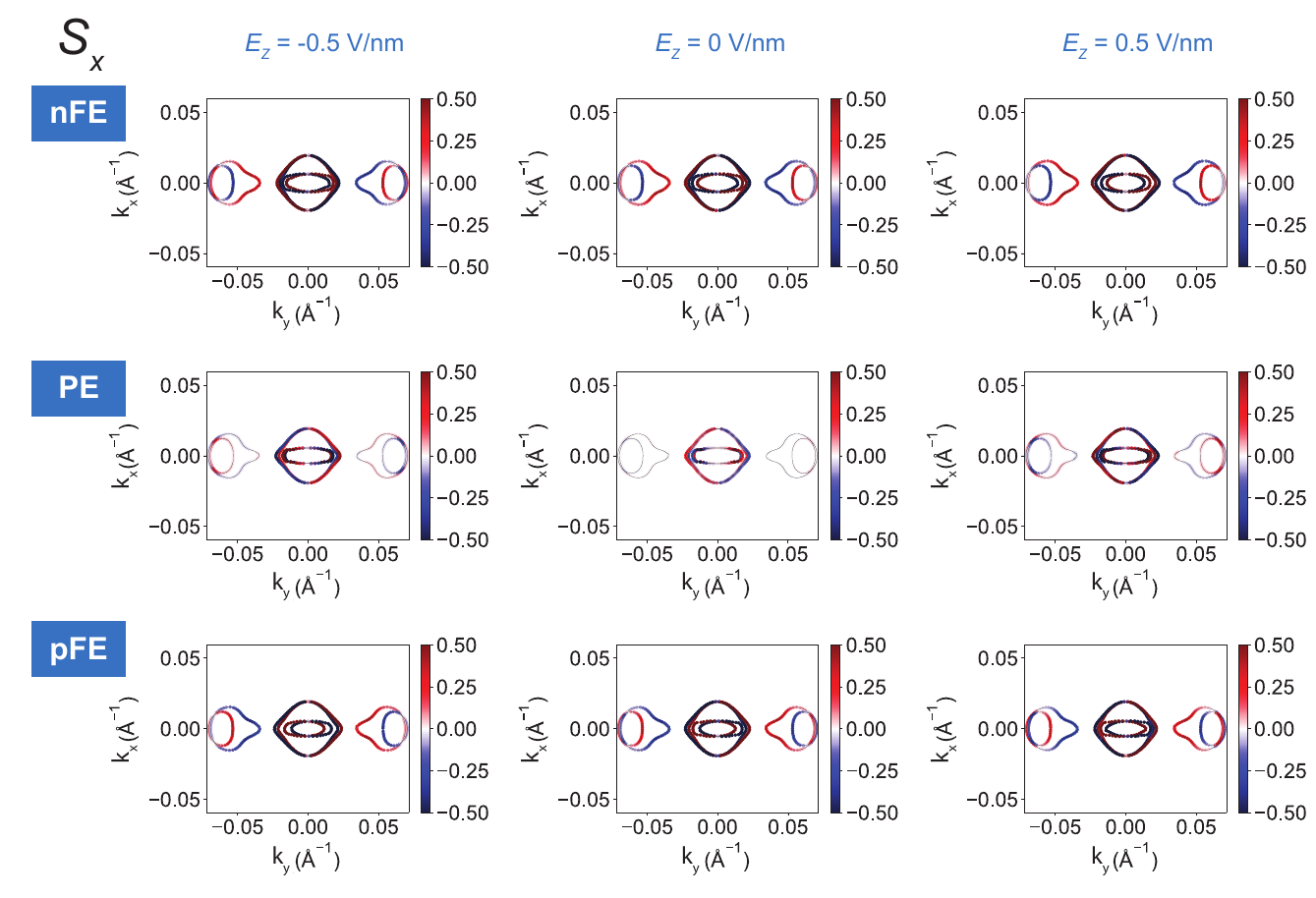}
	\caption{Spin texture projection of bilayer MoTe$_2$ in the ferroelectric phase with negative polarization (nFE), paraelectric phase (PE), and ferroelectric phase with positive polarization (pFE) at $E = E_F - 0.01$ eV for $S_x$ component under the out-of-plane electric field of $E_z = -0.5$, 0, and 0.5 V/nm.}
\label{S8}
\end{figure*}
\clearpage

%%%%%%%%%%%%%%%%%%% S-FIGURE 9 %%%%%%%%%%%%%
%
\begin{figure*}[t] 
    \centering
	\includegraphics[width=\linewidth]{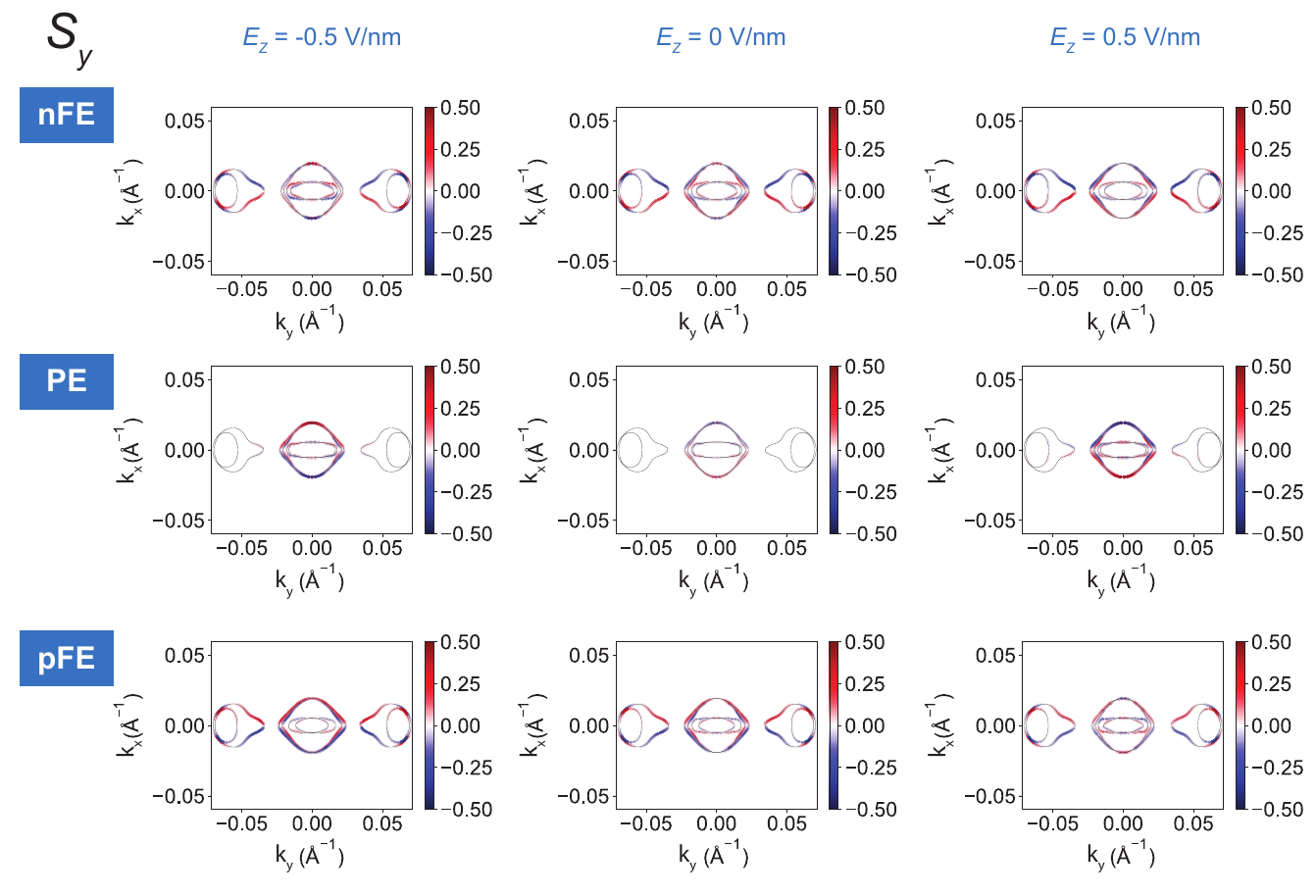}
	\caption{Spin texture projection of bilayer MoTe$_2$ in the ferroelectric phase with negative polarization (nFE), paraelectric phase (PE), and ferroelectric phase with positive polarization (pFE) at $E = E_F - 0.01$ eV for $S_y$ component under varying out-of-plane electric field of $E_z = -0.5$, 0, and 0.5 V/nm.}
\label{S9}
\end{figure*}
\clearpage

%%%%%%%%%%%%%%%%%%% S-FIGURE 10 %%%%%%%%%%%%%
%
\begin{figure*}[t] 
    \centering
	\includegraphics[width=\linewidth]{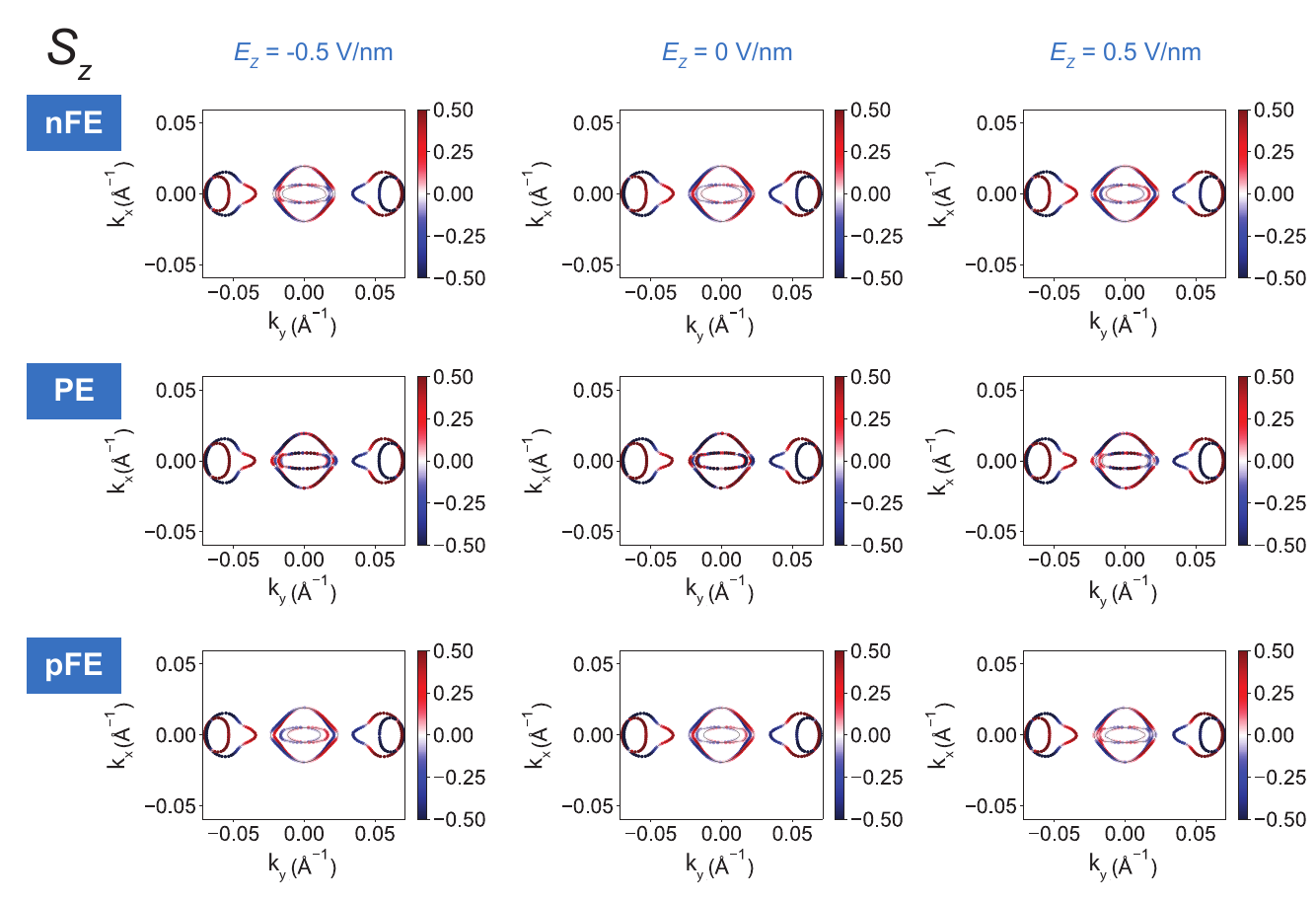}
	\caption{Spin texture projection of bilayer MoTe$_2$ in the ferroelectric phase with negative polarization (nFE), paraelectric phase (PE), and ferroelectric phase with positive polarization (pFE) at $E = E_F - 0.01$ eV for $S_z$ component under varying out-of-plane electric field of $E_z = -0.5$, 0, and 0.5 V/nm.}
\label{S10}
\end{figure*}
\clearpage

%%%%%%%%%%%%%%%%%%% S-FIGURE 11 %%%%%%%%%%%%%
%
\begin{figure*}[t] 
    \centering
	\includegraphics[width=\linewidth]{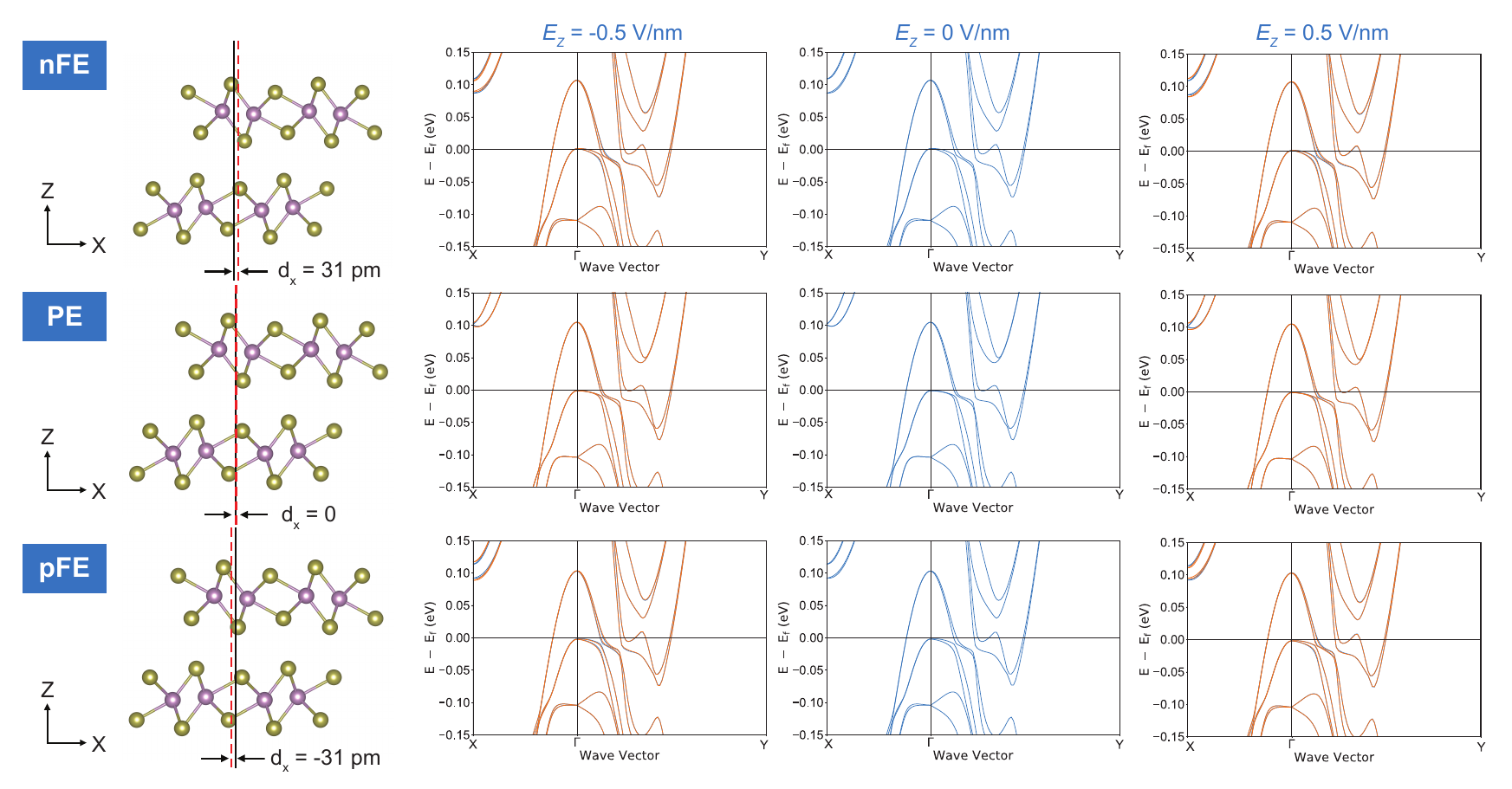}
	\caption{Electronic band structure of bilayer MoTe$_2$ in nFE, PE, and pFE phase under varying out-of-plane electric field of $E_z = -0.5$, 0, and 0.5 V/nm.}
\label{S11}
\end{figure*}
\clearpage

%%%%%%%%%%%%%%%%%%% S-FIGURE 12 %%%%%%%%%%%%%%
%
\begin{figure*}[t] 
    \centering
	\includegraphics[width=.4\linewidth]{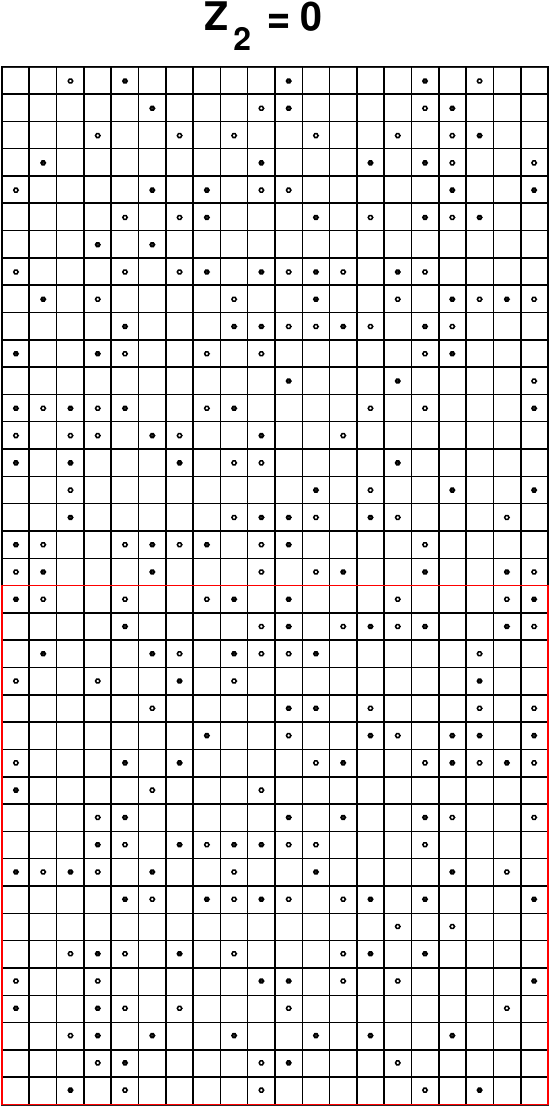}
	\caption{Z$_2$ topological invariant. Z$_2$ topological invariant was calculated for bilayer MoTe$_2$ using the $n$-field method~\cite{Fukui2007nfield}. The result shows bilayer MoTe$_2$ is a Z$_2$-trivial system.}
\label{S12}
\end{figure*}
\clearpage
%%%%%%%%%%%%%%%%%% REFERENCES %%%%%%%%%%%%%%%%%%%%%%
\clearpage
\section*{References}
\bibliographystyle{apsrev4-2}
\bibliography{Supplementary}